\documentclass[prb,twocolumn,superscriptaddress,showpacs,amsmath,amssymb]{revtex4}

\usepackage{graphicx}
\usepackage{latexsym}
\usepackage{amsmath}
\usepackage{amssymb}
\usepackage{amsfonts}
\usepackage{color}
\usepackage{bm}
\usepackage{verbatim}

\begin{document}
\title{
Electronic structure and heat transport in multivortex configurations in
mesoscopic superconductors}

\author{A.~S.~Mel'nikov, D.~A.~Ryzhov, M.~A.~Silaev}
\affiliation{Institute for Physics of Microstructures, Russian
Academy of Sciences, 603950 Nizhny Novgorod, GSP-105, Russia}

\date{\today}

\begin{abstract}
On the basis of the Bogoliubov--de Gennes theory we study the
transformation of the quasiparticle spectrum in the mixed state of a
mesoscopic superconductor, governed by an external magnetic field. We
analyze the low energy part of the excitation spectrum and investigate the
field--dependent behavior of anomalous spectral branches crossing the
Fermi level. Generalizing the Caroli--de Gennes--Matricon approach we
present an analytical solution describing the anomalous branches in a
vortex with an arbitrary winding number. We also study the spectrum
transformation caused by the splitting of a multiquantum vortex into a set
of well separated vortices focusing mainly on a generic example of a
two--vortex system. For vortices positioned rather close to the sample
surface we investigate the effect of the quasiparticle reflection at the
boundary on the spectrum and the density of states at the Fermi level.
Considering an arbitrary surface curvature we study the disappearance of
an anomalous spectral branch for a vortex leaving the sample. The changes
in the vortex configuration and resulting transformation of the anomalous
branches are shown to affect strongly the density of states and the heat
conductance along the magnetic field direction.
\end{abstract}


\maketitle

\section{Introduction}

Modern technology development provides a unique possibility to study
exotic vortex states in mesoscopic superconducting samples of the size of
several coherence lengths~\cite{Mesovortices, Meso-Diagram, Geim}. Tuning
the external applied magnetic field one can switch between a rich variety
of energetically favorable or metastable vortex configurations which can
not be realized in bulk systems. Of particular interest is a possibility
to obtain multiquanta (giant) vortex states with winding numbers larger
than unity for certain intervals of external magnetic field (see, e.g.,
Ref. \onlinecite{Meso-Diagram}). The merging of individual vortices into a
multiquantum one occurs under the influence of screening currents which
push the vortices to the sample center. Note that alternatively the stable
multiquanta vortices can appear  even in a bulk superconductor because of
the pinning on columnar defects with radii of the order of the coherence
length~\cite{M-quantum-defect}. Experimentally the vortex configurations
in mesoscopic systems and the phase transitions between them can be
studied, e.g., by Hall probe measurements of the branching of the
magnetization curve~\cite{Geim} or by observation of the vortex entry into
the sample using the point contact techniques~\cite{Peeters}.

Changing the number and arrangement of the flux lines we can tune the low
energy excitation spectrum which is known to be responsible for low
temperature thermodynamic and transport properties of the sample. The
mechanism of such changes in the subgap quasiparticle spectrum is
associated with the modification of the anomalous energy branches crossing
the Fermi level. For well separated vortices positioned at distances much
larger than the core radius the behavior of the anomalous branches can be
described by the Caroli--de Gennes--Matricon (CdGM) theory~\cite{CdGM}.
For each individual vortex the energy $\varepsilon(\mu)$ of a subgap state
varies from $-\Delta_0$ to $+\Delta_0$ as  one changes the angular
momentum $\mu$ defined with respect to the vortex axis. At small energies
$|\varepsilon|\ll\Delta_0$ the spectrum is a linear function of $\mu$:
$\varepsilon(\mu)\simeq-\mu\omega$, where
$\omega\approx\Delta_0/(k_\perp\xi)$, $\Delta_0$ is the superconducting
gap value far from the vortex axis, $k_\perp=\sqrt{k_F^2-k_z^2}$, $k_F$ is
the Fermi momentum, $k_z$ is the momentum projection on the vortex axis,
$\xi=\hbar V_F/\Delta_0$ is the coherence length, $V_F$ is the Fermi
velocity, and $\mu$ is half an odd integer.

With the decrease in the intervortex distance the quasiparticle tunneling
between the vortex cores comes into play resulting in the modification of
the anomalous branches \cite{Melnikov-Silaev-2006}. Finally, when the
vortex cores merge one obtains a multiquantum vortex with a certain
winding number $M$. The number of anomalous branches per spin
projection~\cite{Volovik-1993} is conserved during this process of
crossover from $M$ individual flux lines to the $M$--quantized giant
vortex and equals to the vorticity $M$. Previously, the behavior of the
anomalous branches in a multiquantum vortex has been investigated
numerically~\cite{multi-spectrum-num} and analytically for a  step--like
model profile of the order parameter in the
core~\cite{Melnikov-Vinokur-2002}. For vortices with an even vorticity all
the anomalous branches cross the Fermi level at nonzero impact parameters
$b=-\mu/k_\perp$:
\begin{equation}
\label{Volovik-spectr}
  \varepsilon(\mu)\sim-(\mu\pm\mu_{j})\Delta_0/(k_\perp\xi)\,
\end{equation}
where $j=1 ... M/2$, $\mu_{M/2}\sim k_\perp\xi$. For a vortex with an odd
winding number there appears a branch crossing the Fermi level at zero
impact parameter.

The wave functions of the subgap states are localized inside the vortex
core because of the Andreev reflection of quasiparticles at the core
boundary. Any additional normal scattering process should modify the
behavior of the anomalous spectral branch. Such modification can be caused
even by atomic size impurities, as it was predicted by Larkin and
Ovchinnikov in Ref. \onlinecite{Larkin-Ovchinnikov-1998}. For a vortex
approaching a flat sample boundary the distortion of the local density of
states (DOS) profile has been analyzed in Ref. \onlinecite{dahm}
numerically on the basis of Eilenberger theory both for $s$-- and $d$--
wave pairing symmetries. Certainly the role of normal scattering at the
boundaries can be of particular importance for vortices trapped in
mesoscopic samples. For a single vortex placed in a superconducting
cylinder an appropriate spectrum transformation has been studied in Refs.
\onlinecite{PRL-2005, PRB-2007}.

Experimentally the behavior of the anomalous branches can be probed, e.g.,
by the scanning tunneling microscopy (STM) or by the heat transport
measurements. The modern STM techniques is a unique tool for the study of
the local DOS profiles and, thus, could provide us the information about
the number and configuration of the spectral branches crossing the Fermi
level. An important advantage of the heat conductance measurements along
the vortex lines is associated with the fact that probing the number and
transparency of quasiparticle transport channels this method appears to be
sensitive also to the $k_z$--dispersion of the spectrum and, in
particular, to the group velocity of quasiparticle modes propagating along
the vortex cores~\cite{PRB-2007, KMV03}. Indeed, it is the small group
velocity of CdGM states which is responsible for a strong suppression of
the heat conductance $\kappa_v\sim T^2k_F\xi/(\hbar\Delta_0)$ along a
singly--quantized vortex core as compared to the Sharvin's conductance
$\kappa_{Sh}\sim T (k_F\xi)^2/\hbar$ of a normal metal wire of the radius
$\xi$ at a certain temperature $T$:
\begin{equation}
\label{NL-est}
  \frac{\kappa_v}{\kappa_{Sh}}
  \sim\frac{1}{k_F\xi}\frac{T}{\Delta_0}\ll 1 \ .
\end{equation}
Within the Landauer approach such suppression of the vortex heat
conductance can be understood as a consequence of a strong reduction of
the effective number of conducting modes $N_v=\kappa_v/\kappa_0$, where
$\kappa_0=\pi T/(3\hbar)$ is the universal heat conductance per conducting
mode in a normal metal. Taking the interlevel spacing
$\omega_0\approx\Delta_0/(k_F\xi)$ for the anomalous branch at $k_z=0$,
one obtains $N_v\sim T/\omega_0$ which agrees with the above
estimation~(\ref{NL-est}). Both the intervortex quasiparticle tunneling
and the normal scattering at the sample boundary are expected to affect
the effective number of conducting modes resulting in the dependence of
the heat transport on the vortex configuration in a mesoscopic
superconductor. The increase in the heat conductance stimulated by the
boundary effects was demonstrated in Ref. \onlinecite{PRB-2007} for a
single--vortex state in a cylinder.

It is the goal of the present paper to study both the transformation of
the anomalous branches and distinctive features of the DOS and heat
transport in different multi-vortex configurations in mesoscopic samples.
We include in our consideration both the spectrum transformation caused by
the giant vortex splitting and the process of an anomalous branch
formation (disappearance) which occurs when a vortex enters (exits) the
sample. In the present study we address only the case of homogeneous
mesoscopic superconductors without any defects or pinning centers.

The paper is organized as follows. To elucidate our main results we start
from a qualitative discussion of the behavior of the anomalous branches in
a mesoscopic superconductor (see Sec. \ref{SEC:picture}). In Sec.
\ref{andreev} we introduce the basic equations used for the spectrum
calculation. In Sec. \ref{SEC:M-vortex} we consider an analytical solution
describing the spectrum of a multiquantum vortex line. In Sec.
\ref{SEC:molecule} we study a generic example of the spectrum
transformation caused by the decay of giant vortices, i.e. the splitting
of a doubly--quantized vortex into two individual singly--quantized
vortices. The influence of the normal reflection at the sample surface on
the quasiparticle spectrum for a vortex positioned close to the boundary
is analyzed in Sec. \ref{SEC:boundary}. In Sec. \ref{SEC:DOS} and Sec.
\ref{SEC:heat-cond} we calculate the density of states  and the thermal
conductance, respectively, using the quasiparticle spectra found in the
previous sections. We summarize our results in
Section~\ref{DiscusSection}. Some of the details of our calculations are
given in appendices.

\section{Transformation of anomalous spectral branches:
qualitative physical picture}
 \label{SEC:picture}

As we discussed in Introduction there are two basic mechanisms responsible
for the transformation of anomalous branches in a multivortex
configuration: (i) the tunneling of quasiparticles between the vortex
cores; (ii) quasiparticle scattering at the sample boundaries which comes
into play when the vortices approach the superconductor surface. To
clarify the key physical consequences of these mechanisms hereafter we
consider two model problems: (i) electronic structure of a multivortex
system positioned rather far from the boundary; (ii) electronic structure
of an individual vortex approaching the sample surface.

\subsection{Effect of intervortex quasiparticle tunneling}

Let us start with a qualitative analysis of the behavior of the anomalous
branches and consider a set of vortex lines parallel to the $z$--axis. It
is the case of intermediate values of magnetic field when vortices are
quite far from the boundary but do not merge into a multiquantum vortex.
In the $(xy)$--plane the vortex centers defined as points of the order
parameter phase singularities (and hence as zeros of the superconducting
order parameter) are positioned at certain coordinates ${\bf r}_i$.

The system is homogeneous in the $z$--direction and, as a result, the
momentum $k_z$ appears to be conserved. The two--dimensional quantum
mechanical problem in the $(xy)$--plane can be strongly simplified
provided the wavelength $k_\perp^{-1}$ is much less than the
superconducting coherence length $\xi$. Thus, following standard
quasiclassical procedure (see Section \ref{andreev} for details) one can
describe the quantum mechanics of quasiparticles using the geometrical
optics picture. An important distinctive feature of this picture in
superconductors is that all the classical trajectories can be
approximately considered as straight lines. The bending of these straight
trajectories is negligible because of a small momentum change $\delta
k\sim 1/\xi$ during the process of quasiparticle scattering at the
inhomogeneous superconducting gap profile, i.e. during the Andreev
reflection. We also can neglect the trajectory bending  caused by magnetic
field, since we assume the cyclotron radius $r_H\sim V_F/\omega_H\sim
k_F\xi^2(H_{c2}/H)$ to exceed all the relevant length scales of our
problem. Here $\omega_H=|e|H/(mc)$ is the cyclotron frequency, $m$ is the
electron effective mass and $H_{c2}$ is the upper critical field.

For quasiparticles propagating along the classical trajectories parallel
to ${\bf k}_\perp =k_\perp(\cos\theta_p,\sin\theta_p)$ we introduce the
angular momenta $\mu=[{\bf r},{\bf k}_\perp]\cdot{\bf z_0}=k_\perp
r\sin(\theta_p-\theta)$ and $\tilde\mu_i=\mu-[{\bf  r}_i,{\bf
k}_\perp]\cdot{\bf z}_0$ defined with respect to the $z$--axis passing
through the origin and with respect to the $i$-th vortex axis passing
through the point ${\bf r}_i$, correspondingly [$(r,\theta,z)$ is the
cylindrical coordinate system]. Neglecting the quasiparticle tunneling
between the vortex cores and the normal scattering at the sample boundary
we get degenerate CdGM energy branches: $\varepsilon_i=-\omega\tilde\mu_i$
for $|\varepsilon|\ll\Delta_0$. For a fixed energy $\varepsilon$ we can
define a set of crossing quasiclassical orbits in the plane $(\mu
,\theta_p)$: $\mu_i(\theta_p) =-\varepsilon/\omega +[{\bf r}_i,{\bf
k}_\perp]\cdot{\bf z}_0$.

\begin{figure}[htb]
\centerline{\includegraphics[width=75mm]{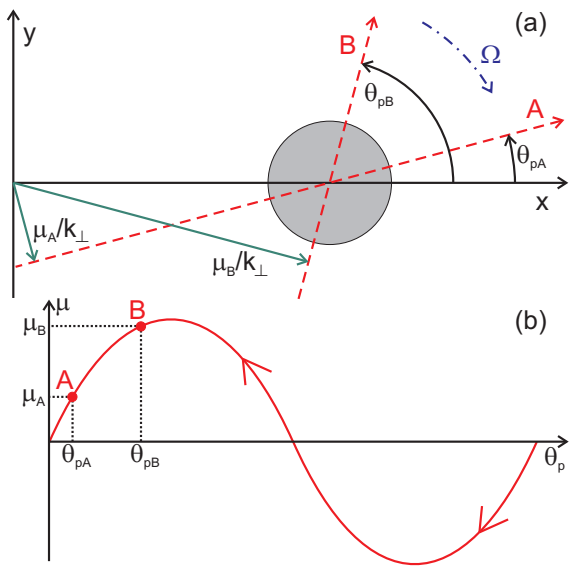}}
 \caption{\label{FIG:RealSpaceTraj}
Schematic plot of trajectory precession around vortex line in real space
(a) and corresponding quasiclassical orbit for $\varepsilon=0$ in the
$(\mu,\theta_p)$--plane (b). Vortex core is shown by grey circle.}
\end{figure}

\begin{figure}[htb]
\centerline{\includegraphics[width=85mm]{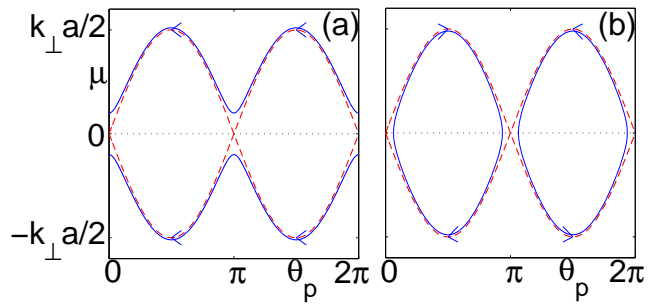}}
 \caption{\label{fig1}
Schematic plots of quasiclassical orbits for $\varepsilon=0$ in the
$(\mu,\theta_p)$--plane (solid lines) for (a) two vortices with
intervortex distance $a$ and (b) vortex near the flat surface ($a/2$ is
the distance between vortex and surface). The orbits for two
non--interacting vortices are shown by dash lines.}
\end{figure}

The quasiclassical orbit in the $(\mu ,\theta_p)$--plane for a single
Abrikosov vortex is shown in Fig. \ref{FIG:RealSpaceTraj}(b).
 %
 %
Each point at this orbit corresponds to a straight trajectory passing
through the  vortex core [Fig. \ref{FIG:RealSpaceTraj}(a)]. Precession of
the quasiclassical trajectory is described by the Hamilton equation:
$\hbar\partial{\theta_p}/\partial t=\partial\varepsilon/\partial\mu$ which
provides us the precession frequency $\Omega=-\omega/\hbar$. This
precession is a result of the small deviation from the exact Andreev
backscattering of quasiparticles in the vortex core. In Fig.
\ref{FIG:RealSpaceTraj}(b) the direction of the trajectory precession is
shown by arrows. The discrete spectrum of subgap quasiparticle states can
be found using the Bohr--Sommerfeld rule. In the case of several vortices
we have several crossing quasiclassical orbits in the
$(\mu,\theta_p)$--plane. These orbits are shown by dash lines in Fig.
\ref{fig1}(a) for a particular case of two vortices with ${\bf
r}_1=(-a/2,0)$ and ${\bf r}_2=(a/2,0)$.
 %
 %
Each crossing point of quasiclassical orbits $\mu_i(\theta_p)$ and
$\mu_j(\theta_p)$ corresponds to the trajectories passing through
the cores of $i-$th and $j-$th vortices. It is natural to expect
that the degeneracy at these points will be removed if we take
account of a finite probability of quasiparticle tunneling between
the cores. Let us consider the vicinity of the degeneracy point,
e.g., $\theta_p=0$ [see Fig. \ref{fig1}(a)]. The trajectory
characterized by the angle $|\theta_p|\ll \xi/a$ passes through
both vortex cores, and therefore the wave function along such
trajectories can be found as a superposition of two states
localized at different vortices and having close energies:
$\varepsilon_{v1}=-\omega[\mu-(k_\perp a/2)\sin\theta_p]$ and
$\varepsilon_{v2}=-\omega[\mu+(k_\perp a/2)\sin\theta_p]$. The
transformation of the quasiclassical spectrum occurs due to
overlapping of the corresponding wave functions can be described
using a standard quantum mechanical approach  describing a
two--level system~\cite{LL-III}, which yields the secular equation
 \begin{equation}
 \label{secularVV}
 (\varepsilon-\varepsilon_{v1})(\varepsilon-\varepsilon_{v2})=(\delta\varepsilon)^2,
 \end{equation}
and resulting splitting of isoenergetic lines near the degeneracy point
($\theta_p=0$ for our example):
\begin{equation}
\label{orbit1}
  \varepsilon=-\omega\mu\pm\sqrt{\omega^2(k_\perp a/2)^2\theta_p^2+(\delta
  \varepsilon)^2}\,.
\end{equation}
The tunneling of quasiparticles between vortex cores is determined by the
exponentially small overlapping of wave functions localized near the cores
and results in the splitting of energy levels:
$\delta\varepsilon\sim\Delta_0\exp(-k_F a_{ij}/(k_\perp\xi))$, where
$a_{ij}=|{\bf r}_i-{\bf r}_j|$ is the distance between vortex lines, and
$k_F a_{ij}/k_\perp$ is the distance between vortex centers along the
trajectory. The estimate for the splitting
$\delta\mu\simeq\delta\varepsilon/\omega$ of isoenergetic lines in the
$(\mu,\theta_p)$--plane reads [see Eq. (\ref{orbit1})]:
\begin{equation}
\label{delta-mu}
  \delta \mu(a_{ij})\sim
  k_\perp\xi\exp\left(-\frac{k_F a_{ij}}{k_\perp\xi}\right)\,.
\end{equation}
As a result, we get the orbits $\mu^*_i(\theta_p)$ with a qualitatively
new behavior [solid lines in Fig. \ref{fig1}(a)]: each of these orbits
consists of parts corresponding to the classical quasiparticle
trajectories passing through the cores of different vortices.

The tunneling between the cores of the different vortices becomes
significant when the energy splitting $\delta \varepsilon$ is comparable
with the interlevel spacing $\omega_0$, i.e., when $\delta\mu\gtrsim 1$ in
Eq. (\ref{delta-mu}). According to the above condition on
$\delta\mu(a_{ij})$ the tunneling is most efficient for $k_\perp=k_F$ and
$a_{ij}<a_c$, where $a_c\simeq\xi\ln(k_F\xi)$ is a critical intervortex
distance. Using the percolation theory language we can consider the
vortices to be bonded if $a_{ij}<a_c$ and define a cluster in a flux line
system as a set of $M$ vortices bonded either directly or via other
vortices. Certainly in mesoscopic superconductors the cluster dimensions
$L_v$ can not exceed the sample size. The cluster is characterized by a
set of hybridized quasiparticle states: with a change in the ${\bf
k}_\perp$--direction the wave function experiences a number of subsequent
transitions between the cores of neighboring vortices. Taking, e.g., the
upper quasiclassical orbit in Fig. \ref{fig1}(a) we obtain the wave
function concentrated near the cores of the right and left vortices for
the angular intervals $0<\theta_p<\pi$ and $\pi<\theta_p<2\pi$,
respectively. Further decrease in the intervortex distance results in the
increase in the tunneling probability and, thus, the increase in
$\delta\mu (a_{ij})$. Finally, for $a_{ij}\rightarrow 0$ we get a set of
$M$ lines $\mu ={\rm const}$ parallel to the $\theta_p$ axis, i.e., $M$
anomalous branches crossing zero energy at angular independent impact
parameters and corresponding to the $M$--quantum vortex. Certainly this
limit can be realized only in mesoscopic samples.

Within the quasiclassical approach one can  estimate the intervortex
tunneling efficiency using the Landau--Zehner transition theory. Let us
consider the vicinity of the degeneracy point, e.g., $\theta_p=0$ [see
Fig. \ref{fig1}(a)]. The tunneling probability of transition from one
quasiclassical orbit to another is given by the expression~\cite{LL-III}
\begin{equation}
\label{LZ}
  W=\exp\left(-4{\rm Im}\int_0^{i\theta_p^*}\mu(\theta_p)
  \,d\theta_p\right)\,,
\end{equation}
where $\theta_p^*=2\delta\varepsilon/(\omega k_\perp a)$ and
$\mu(\theta_p)$ should be taken from the Eq. (\ref{orbit1}) with lower
sign. Finally, we obtain the following estimate for the tunneling
probability:
 \begin{equation}\label{LZest1}
  W=\exp(-2\pi(\delta\mu/\Delta\mu)^2) \ ,
 \end{equation}
where $\Delta\mu=\sqrt{k_\perp a}$ is the quantum mechanical uncertainty
of the angular momentum. Therefore, we can neglect the tunneling between
quasiclassical orbits while $\delta\mu\gtrsim\Delta\mu$.

Following Ref. \onlinecite{Kopnin-Volovik-1996} one can obtain the
discrete energy levels applying the Bohr--Sommerfeld quantization rule for
canonically conjugate variables $\mu$ and $\theta_p$:
\begin{equation}
\label{bohr}
  \int_0^{2\pi n_\theta}\mu(\theta_p) d\theta_p=2\pi(n+\beta),
\end{equation}
where $n$ and $n_\theta$ are integers, $2\pi n_\theta$ is the period of
the $\mu(\theta_p)$ function ($1\leq n_\theta\leq M$), and $\beta$ is of
the order unity. The period of  $\mu(\theta_p)$ can be larger than $2\pi$
($n_\theta > 1 $) if the Landau--Zehner transitions between quasiclassical
orbits are not negligible. Depending on the ratio $\delta\mu/\Delta\mu$
one should apply this quantization rule either to the orbits
$\mu_i(\theta_p)$ or to the orbits $\mu^*_i(\theta_p)$. In the momentum
region
\begin{equation}
  k_F\sqrt{1-[\min(a_{ij})/a_c]^2}\ll|k_z|<k_F
\end{equation}
we can neglect the splitting of isoenergetic lines [$\delta\mu\ll
\Delta\mu$] and Eq. (\ref{bohr}) written for the orbits $\mu_i
(\theta_p)$ gives us the CdGM spectrum with a minigap
$\omega_0/2=\omega (k_z =0)/2$. For ${\rm min}(a_{ij})>a_c$  the
CdGM expression holds for the entire momentum range. For vortices
forming a cluster the quasiparticle states bonded by intervortex
tunneling appear in a finite momentum interval $|k_z|<k_z^*$,
where
\begin{equation}
\label{Threshold}
  k_z^*=k_F\sqrt{1-[\min(a_{ij})/a_c]^2}\,.
\end{equation}
In this limit the quasiparticle tunneling between the cores results in the
qualitative modification of spectrum which can be obtained by substituting
$\mu^*_i(\theta_p)$ into Eq. (\ref{bohr}):
\begin{equation}
\label{double}
  \varepsilon_{ni}(k_z)\approx\frac{\Delta_0}{\xi}
  \left[\frac{n+\beta}{k_\perp}+b_i ({\bf r}_1,..{\bf r}_M)\right]\,,
\end{equation}
where $i=1 .. M$. The spectrum (\ref{double}) is similar to the
one of a multiquanta vortex \cite{Volovik-1993,
multi-spectrum-num, Melnikov-Vinokur-2002} which recovers in the
limit $a_{ij}\rightarrow 0$ when $|b_i|\lesssim \xi$. The
multi--vortex cluster geometry and its size $L_v$ determine the
effective impact parameters $b_i ({\bf r}_1,..{\bf r}_M)$ which
vary in the range $-L_v\lesssim b_i\lesssim L_v$. Taking a two--
(three--) vortex cluster with $\xi<a<a_c$ as an example we get
$b_{1,2}\sim\pm a$ ($b_{1,3}\sim\pm a$, $b_2=0$). Contrary to the
CdGM  case the spectrum branches (\ref{double}) can cross the
Fermi level as functions of $k_z$ as we decrease the intervortex
distance $a$ and the minigap is suppressed. The DOS consists of
$M$ sets of van Hove singularities corresponding to the extrema of
$\varepsilon_{ni}(k_z)$ branches. The energy interval between the
peaks belonging to each set is $\omega_0$. For a fixed energy the
DOS as a function of $a$ exhibits oscillations with the period of
the order of the atomic length scale. Experimentally the
intervortex distance can be controlled by a varying magnetic
field. For typical values $a\sim\sqrt{\phi_0/H}$ we get the
following field scale of DOS oscillations: $\delta H/H\sim
\sqrt{\hbar\omega_H/\varepsilon_F}$, where $\varepsilon_F$ is the
Fermi energy. The oscillatory behavior should affect both
thermodynamic and transport properties at low temperatures though
in real experimental conditions the DOS peak structure is
certainly smeared due to the various mechanisms of level
broadening, e.g., finite temperature, fluctuations in vortex
positions, impurity scattering effects, etc. It should be noted
that for typical values $k_F\xi=10^2 - 10^3$ the critical distance
$a_c/\xi\sim 4-6$ exceeds the core radius and the spectrum
transformation starts at the fields $H\sim \phi_0/a_c^2\sim
H_{c2}[{\rm \ln}(k_F\xi)]^{-2}$ when the vortices are indeed
well--separated.

\subsection{Effect of normal reflection of quasiparticles at the boundary}
\label{reflection}

The above discussion of the behavior of anomalous branches assumed
a negligible role of the normal scattering of quasiparticles at
the sample boundary. Such assumption is certainly no more valid
when the intervortex distance becomes so large that some of the
individual vortices approach the sample boundary and their
spectrum is determined by the interplay of Andreev reflection and
normal scattering at the boundary. For rather small samples this
interplay can influence the spectrum even for a vortex positioned
at the sample center \cite{PRL-2005, PRB-2007}.

\begin{figure}[htb]
\centerline{\includegraphics[width=1.0\linewidth]{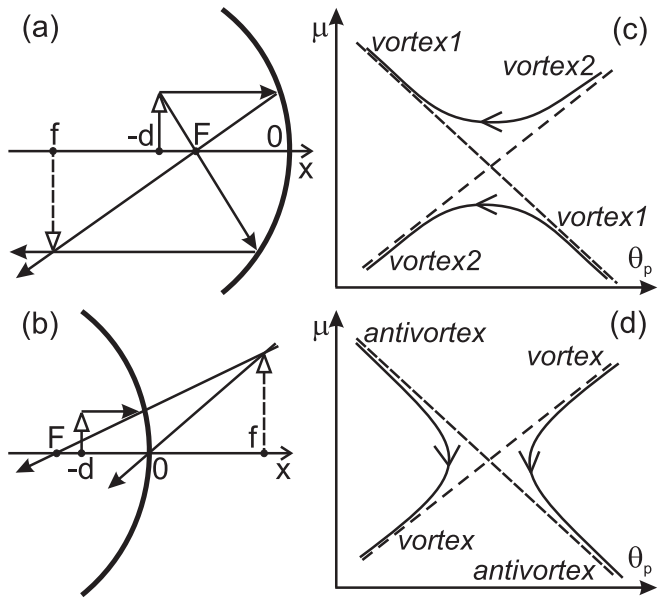}}
  \caption{\label{paraboloid}
The geometrical optics analogy for the reflection of quasiclassical
trajectories at the concave ($F<0$) parabolic boundary (left panel) and
the corresponding modification of the isoenergetic lines (right panel) for
$d>-F$ (a), (c), and for $d<-F$ (b), (d).}
\end{figure}

In order to focus on the role of the boundary effects we consider a model
situation when the intervortex quasiparticle tunneling can be neglected
(i.e. $a_{ij}>a_c$) because of the rather large cluster size. Thus, we can
study the spectrum modification for a single vortex approaching the
boundary characterized by a certain curvature in the plane perpendicular
to the vortex axis (see Fig. \ref{paraboloid}).
 %
 %
For the sake of simplicity we restrict ourselves to the case of a smooth
and specularly reflecting surface. Obviously the quasiclassical spectrum
should be disturbed most strongly for trajectories which return back to
the vortex core after the normal reflection at the boundary. These
trajectories experience the reflection from a rather narrow surface region
near the point positioned at the surface at a minimal distance $d$ from
the vortex center. In this region we may consider the surface profile as a
parabolic cylinder with a certain focal distance $F$. Introducing a polar
coordinate system $(r,\theta)$ with the origin in the vortex center one
obtains the following equation describing this parabolic cylinder at small
$\theta$ angles: $r(\theta)=d\{1+[1/2+d/(4F)]\theta^2\}$. Note that we
should put $d/F>-2$ so that the distance $d$ is indeed a minimal distance
to the surface. As it is shown in Fig. \ref{paraboloid} the vortex center
is positioned at the optical axis of the parabolic mirror. The
trajectories experiencing normal reflection and passing twice through the
core can be considered within the paraxial approximation.

In this case the scattering rules for trajectories reflecting from the
surface can be obtained employing an analogy with a textbook picture
describing the system of rays and images in geometrical optics. For a
particular case of a concave parabolic mirror ($F<0$) the system of rays
is schematically shown in Figs. \ref{paraboloid}(a), (b). For the object
(white solid arrow) situated at the distance $d$ from the mirror the image
(white dash arrow) is formed by the reflected rays at the coordinate
$f=-d/h$, where $h=-(1+d/F)$. The type of image is determined by the sign
of $h$: if $h>0$ the image is real, i.e. it has the coordinate $f<0$ and
is situated at the same side of the mirror relative to the object [see
Fig. \ref{paraboloid}(a)], otherwise the image is virtual with $f>0$ and
situated at the other side of the mirror [see Fig. \ref{paraboloid}(b)].
Using these well--known results the parameters of reflected trajectories
can be derived from the simple trigonometry.

Let us consider the trajectory which makes a small angle $|\theta_p|\ll 1$
with the $x$--axis and has a small impact parameter $|b|\ll d$. Then, the
reflected trajectory has the angle $\tilde{\theta}_p=\pi+h\theta_p$ and
its impact parameter is $\tilde{b}=h b$. The impact parameters of incident
and reflected trajectories are defined relative to the point $\theta=0$,
$r=d$ positioned at the surface at the minimum distance to the vortex
center. This point coincides with the coordinate system origin in Figs.
\ref{paraboloid}(a),(b).

In the $(\mu,\theta_p)$--plane one can define isoenergetic lines
corresponding to the incident and reflected trajectories. The intersection
of these lines occurring for $|\theta_p|\ll 1$ corresponds to the
situation when both the incident and reflected trajectories pass through
the vortex core. The degeneracy at the intersection point should be
removed by the splitting of isoenergetic lines due to the interaction of
vortex core states. One can obtain two qualitatively different regimes of
splitting, determined by the ratio between the focal distance $F$ of the
parabolic mirror and the distance from the vortex to the surface $d$, i.e.
by the sign of $h$. The splitting corresponds to the continuous transition
from one isoenergetic line to another. The velocity of motion along
isoenergetic lines is determined by the angular velocities of trajectory
precession, given by $\Omega=\partial\theta_p/\partial t$ and
$\tilde{\Omega}=\partial\tilde{\theta}_p/\partial t=h\Omega$ for the
incident and reflected trajectories, correspondingly. Therefore, if
$\Omega$ and $\tilde{\Omega}$ have the same signs ($h>0$), then the
directions of precession of the incident and reflected trajectories
coincide. In this case, the orbit transformation is analogous to the one,
that have been obtained for the pair of interacting vortices [Figs.
{\ref{paraboloid}}(c) and \ref{fig1}(a)]. Otherwise, if $h<0$, the
incident and reflected trajectories precess in different directions,
therefore the splitting occurs as shown in Figs. {\ref{paraboloid}}(d) and
\ref{fig1}(b) analogously to the transformation of isoenergetic lines
which can be obtained for the interacting vortex and antivortex.

To study the spectrum transformation let us consider in detail the
simplest configuration: the vortex line is situated at the point
$(-a/2,0)$ near the flat boundary of superconductor, occupying the
half--space $x<0$.  If we neglect the normal reflection of quasiparticles
at the boundary, the isoenergetic line in the $(\mu,\theta_p)$--space is
given by $\mu_v(\theta_p)=-\varepsilon/\omega-(k_\perp a/2)\sin\theta_p$,
shown for the particular case $\varepsilon=0$ by the dash line in Fig.
\ref{fig1}(b). The flat boundary is characterized by the infinite focal
distance: $F=\infty$, therefore $h=-1$ and we obtain the mapping of
incident and reflected trajectories according to the following rule:
$\tilde{b}=-b $ and $\tilde{\theta}_p=\pi-\theta_p$. Then, we obtain
another isoenergetic line:
$\mu_{av}(\theta_p)=-\mu_v(\pi-\theta_p)=\varepsilon/\omega+(k_\perp
a/2)\sin\theta_p$, which corresponds to the reflected trajectories and is
shown in Fig. \ref{fig1}(b) by another dash curve. Note that the
isoenergetic line $\mu_{av}(\theta_p)$ (with the opposite direction of
trajectory precession) coincides with the isoenergetic line corresponding
to an antivortex placed at the point $(0,a/2)$ outside the superconductor.
It means that the spectrum of the vortex near the flat surface can be
obtained considering the spectrum of the vortex--antivortex system.
Indeed, the pair potential of the vortex--antivortex system is invariant
under the reflection at the $x=0$ plane: $\Delta(x,y)=\Delta(-x,y)$,
yielding the symmetry of the quasiparticle wave function:
$\hat\Psi(x,y)=\pm \hat\Psi(-x,y)$. The odd wave functions obey the
boundary condition $\hat\Psi(0,y)=0$. The even wave functions obey the
boundary conditions $\partial\hat\Psi(0,y)/\partial x=0$ and the
corresponding energy levels should be omitted in order to obtain the
spectrum of the vortex near the surface. The isoenergetic lines
$\mu_{v,av}(\theta_p)=\pm[\varepsilon/\omega+(k_\perp a/2)\sin\theta_p]$
intersect at certain points, e.g., at $\theta_p=\pi n$ for
$\varepsilon=0$, where $n$ is integer. The degeneracy at these points is
removed by the splitting of isoenergetic lines, shown in Fig.
\ref{fig1}(b) by the solid lines. Considering the interaction of the two
quantum states with close energies $\varepsilon_v=-\omega[\mu+(k_\perp
a/2)\sin\theta_p]$ and $\varepsilon_{av}=\omega[\mu-(k_\perp
a/2)\sin\theta_p]$, we obtain again the secular equation (\ref{secularVV})
with $\varepsilon_{v1}=\varepsilon_v$ and
$\varepsilon_{v2}=\varepsilon_{av}$. Therefore, the quasiclassical orbits
near the degeneracy point $\theta_p=0$ are given in this case by the
following expression:
\begin{equation}
\label{orbit2}
  \varepsilon=\pm\sqrt{(\omega\mu)^2+(\delta \varepsilon)^2}
  -\omega(k_\perp a/2)\theta_p\,.
\end{equation}
The classically forbidden angular domain at $\varepsilon=0$ has the width
$\delta\theta_p=4\delta\varepsilon/(\omega k_\perp a)$. One can assume
that the appearance of such classically forbidden domain explains the deep
structure in the local DOS profile observed numerically in Ref.
\onlinecite{dahm} for a vortex near the flat boundary of an $s$--wave
superconductor. As we show below, the classically forbidden angular domain
results in the suppression of the overall DOS and we propose that this
mechanism should be responsible for the anomalous spectrum branch
disappearance when the vortex exits the sample.

To estimate the tunneling probability between the quasiclassical orbits we
again apply the theory of Landau--Zehner transitions. The Landau--Zehner
tunneling probability is expressed as follows:
\begin{equation}
\label{LZ1}
  W=\exp\left(-4{\rm Im}\int_0^{i\mu^*}\theta_p(\mu)\,d\mu\right)\,,
\end{equation}
where $\mu^*=\delta\varepsilon/\omega$ and $\theta_{p}(\mu)$ should be
taken from Eq. (\ref{orbit2}) with the upper sign. Finally, we obtain the
tunneling probability as
$W\sim\exp(-2\pi(\delta\theta_p/\Delta\theta_p)^2)$, where
$\Delta\theta_p\sim(k_\perp a)^{-1/2}$ is the quantum mechanical
uncertainty of the trajectory orientation angle. Thus, we can neglect
Landau--Zehner effects while $\delta\theta_p\gtrsim\Delta\theta_p$, i.e.,
for $a<a_c$, where $a_c\sim\xi\ln (k_\perp\xi)$ is the critical distance,
which appears to be the same as for a two--vortex system.

\section{Andreev equations}
\label{andreev}

Our further consideration is based on the Bogoliubov--de Gennes (BdG)
equations for particle-- ($u$) and hole--like ($v$) parts of the wave
function, which have the following form:
\begin{eqnarray}
\label{BdG}
  \nonumber
  -\frac{\hbar^2}{2m}\left(\nabla^2+k_F^2\right)u+\Delta({\bf r})v
  &=&\varepsilon u\,,\\
  \frac{\hbar^2}{2m}\left(\nabla^2+k_F^2\right)v+\Delta^*({\bf r})u&
  =&\varepsilon v\,.
\end{eqnarray}
Here $\Delta({\bf r})$ is the gap function and ${\bf r}=(x,y)$ is a radius
vector in the plane perpendicular to the magnetic field direction. We
assume the system to be homogeneous along the $z$-- axis, thus, the
$k_z$-- projection of the momentum is conserved and the wave function
takes the form:
  $$
  (u,v)=e^{ik_z z}\hat\Psi({\bf r})\ .
  $$
Then, the two--component wave function $\hat\Psi$ in the momentum
representation can be written as follows:
\begin{equation}
\label{momentum}
  \hat\Psi({\bf r})=\frac{1}{(2\pi\hbar)^2}\iint_{-\infty}^{\infty}
  e^{i{\bf p}{\bf r}/\hbar}\hat \psi_{\bf p}\,d^2{\bf p}\,.
\end{equation}
Let us introduce the polar coordinate system in momentum space ${\bf
p}=p\,(\cos\theta_p,\sin\theta_p)=p\,{\bf p}_0$. Then, the coordinate
operator can be written as follows:
  $$
  \hat{\bf r}= i\hbar\frac{\partial}{\partial{\bf p}}
  =i\hbar\left({\bf p}_0\frac{\partial}{\partial p}
  +\frac{i}{p}[{\bf z}_0,{\bf p}_0]\;\hat\mu\right)\,,
  $$
where operator $\hat\mu$ of $z$--projection of angular momentum is given
by the expression:
 \begin{equation}
 \hat\mu = \frac{1}{\hbar} [{\bf r}, {\bf p}]{\bf z}_0 =
 -i\frac{\partial}{\partial \theta_p}\,.
 \end{equation}
Next, we assume that the quasiparticle wave function can be viewed as a
wave packet with momenta absolute values close to $\hbar k_\perp$. This
assumption is valid with very good accuracy in most superconductors, since
the characteristic length scale of envelopes of quasiparticle waves is
determined by the superconducting coherence length $\xi$, which is
typically much larger than the quasiparticle wave length ($k_F\xi\gg 1$).
Therefore, one can put $p=\hbar k_\perp+q$ ($|q|\ll\hbar k_\perp$) and
obtain:
  $$
  \hat{\bf r}=i\hbar{\bf p_0}\frac{\partial}{\partial q}
  +\frac{i}{2k_\perp}\left\{[{\bf z}_0, {\bf p}_0],
  \frac{\partial}{\partial \theta_p}\right\}\,,
  $$
where $\{...\}$ is an anticommutator. Of course, such approximation is
broken for a very small portion of quasiparticles which propagate very
close to the vortex axis [$2\pi/(k_\perp\xi)\gtrsim1$]. Let us now
introduce a Fourier transformation:
\begin{equation}
\label{st}
  \hat\psi_{\bf p}=\frac{1}{k_\perp}\int\limits_{-\infty}^{+\infty}
  \hat\psi(s,\theta_p)e^{-iqs/\hbar}\,ds\,.
\end{equation}
The variable $s$ is a coordinate along a quasiclassical trajectory, which
is a straight line along the direction of the quasiparticle momentum. The
trajectory orientation angle is given by the $\theta_p$ value. The wave
function in the real space can be found from Eqs.
(\ref{momentum},\ref{st}):
\begin{equation}
\label{xy}
  \hat\Psi(r,\theta)
  =\int\limits_0^{2\pi}e^{ik_\perp r\cos(\theta-\theta_p)}
  \hat\psi(r\cos(\theta-\theta_p),\theta_p)\frac{d\theta_p}{2\pi}\,,
\end{equation}
where $(r,\theta,z)$ is a cylindrical coordinate system. The expression
for coordinate operator in ($s,\theta_p$)--representation reads:
  $$
  \hat{\bf r}=s{\bf p}_0+\frac{i}{2k_\perp}\left\{
  [{\bf z}_0,{\bf p}_0],\frac{\partial}{\partial\theta_p}\right\}\,.
  $$

Then, BdG equations (\ref{BdG}) in ($s,\theta_p$)--representation take the
form $\hat H\hat\psi(s,\theta_p)=\varepsilon\hat\psi(s,\theta_p)$ with the
Hamiltonian given by
\begin{equation}
\label{QuasiH}
  \hat H = -i\hat\sigma_z\frac{\hbar^2 k_\perp}{m}
  \frac{\partial}{\partial s}
  +\left(\begin{array}{cc}
    0 & \Delta(\hat{\bf r}) \\
    \Delta^*(\hat{\bf r}) & 0 \\
  \end{array}\right)\,,
\end{equation}
where $\hat\sigma_x$, $\hat\sigma_y$, $\hat\sigma_z$ are the Pauli
matrices.

Note that the gap function operator $\Delta(\hat{\bf r})$ in Eq.
(\ref{QuasiH}) contains a differential operator $\partial/\partial
\theta_p$, therefore the above quasiclassical equations are still rather
complicated partial differential equations. A further simplification can
be obtained considering eikonal approximation for the angular dependence
of wave function:
  $$
  \hat\psi(s,\theta_p) =e^{iS_e(\theta_p)} \hat g (s,\theta_p) \ ,
  $$
where
  $$
  -\frac{1}{k_\perp}\frac{\partial S_e}{\partial\theta_p}=b(\theta_p)
  $$
is an impact parameter of a quasiclassical trajectory. Assuming that the
angular dependence of $\hat g (s,\theta_p)$ is rather slow, one can
neglect its angular derivatives in Eq. (\ref{QuasiH}). The resulting
Andreev equations characterizing the behavior of the wave function along a
trajectory with a certain orientational angle $\theta_p$ and an impact
parameter $b$ read:
\begin{equation}
\label{andreev-eqs}
  -i\hat\sigma_z\frac{\hbar^2k_\perp}{m}
  \frac{\partial\hat g}{\partial s}+\hat\sigma_x {\rm Re}\Delta(x,y)\hat g
  -\hat\sigma_y{\rm Im}\Delta(x,y)\hat g=\varepsilon\hat g\ ,
\end{equation}
where
\begin{eqnarray}
\label{Coord}
  \nonumber
  x &=& s\cos\theta_p-b\sin\theta_p\,,\\
  y &=& s\sin\theta_p+b\cos\theta_p\,.
\end{eqnarray}
Note that the Andreev equations (\ref{andreev-eqs}) can be obtained
directly from the initial BdG equations (\ref{BdG}) if one applies the
coordinate system transformation (\ref{Coord}) and neglects the
second--order derivatives of the wave function.

\section{Quasiparticle spectrum of a multiquantum vortex}
\label{SEC:M-vortex}

We start our quantitative analysis of quasiparticle spectra with the case
of a multiquantum vortex with vorticity $M$.  In this section we
neglect the effect of normal quasiparticle scattering at the sample
boundary and focus on the peculiarities of the spectrum depending on the
vorticity value. We take the gap profile in the form:
\begin{equation}
\label{hat-H-DM}
  \Delta=D_M\left(r\right)e^{iM\theta}\ .
\end{equation}
In $s,\theta_p$ variables one obtains:
\begin{equation}
\label{hat-H-DM1}
  \Delta=D_M\left(\sqrt{s^2+b^2}\right)e^{iM\theta_p}
  \left[\frac{s+ib}{\sqrt{s^2+b^2}}\right]^M \ .
\end{equation}
Due to the cylindrical symmetry  the $\theta_p$-- dependence of the
function $\hat g$ can be excluded using the gauge transformation
\begin{equation}
\label{Gauge-transf}
\hat g =\exp(iM\hat \sigma_z\theta_p/2)\hat f
\end{equation}
and quasiclassical equations (\ref{andreev-eqs}) take the form
\begin{equation}
\label{Quasiclass}
  -i\hat\sigma_z\frac{\hbar^2k_\perp}{m}
  \frac{\partial\hat f}{\partial s}
  +\hat\sigma_x G_R\hat f-\hat\sigma_y G_I\hat f
  =\varepsilon\hat f \ .
\end{equation}
Here we introduce the functions
\begin{eqnarray}
\nonumber
  G_R=D_M\left(\sqrt{s^2+b^2}\right)
  {\rm Re}\left\{\left[\frac{s+ib}{\sqrt{s^2+b^2}}\right]^M\right\}\ ,\\
  \nonumber
  G_I=D_M\left(\sqrt{s^2+b^2}\right)
  {\rm Im}\left\{\left[\frac{s+ib}{\sqrt{s^2+b^2}}\right]^M\right\}\ .
\end{eqnarray}
To apply the method analogous to the one used in Ref.
\onlinecite{Volovik-1993} for a singly--quantized vortex we note that the
exact solutions of the above equations corresponding to $\varepsilon=0$
can be found in case $G_I\equiv0$:
\begin{equation}
\label{sol0}
  \hat f_\pm=(1,\pm i)\exp\left(\pm\frac{m}{\hbar^2 k_\perp}
  \int_0^s G_R ds\right)\ .
\end{equation}
Provided $G_R$ is an odd function of $s$, which tends to a certain nonzero
value for $|s|\rightarrow\infty$ one of these solutions appears to decay
both at negative and positive $s$ and, thus, we get a localized wave
function corresponding to a midgap bound state. Using this localized
solution as a zero--order approximation for the wave function the spectrum
can be found within the first order perturbation theory assuming that
$|\varepsilon|\ll \Delta_0$.

For an arbitrary value of vorticity the function $G_R$ is not necessary
odd. In order to use the perturbation method described above we apply a
gauge transformation
\begin{equation}
\label{Gauge-transform}
  \hat f =\left(\frac{s+i\hat\sigma_z b}{\sqrt{s^2+b^2}}\right)^\alpha
  \hat w\ ,
\end{equation}
so that new wave function $\hat w$ satisfies the following equation:
\begin{equation}
\label{Quasiclass-Doppler}
  \left[-i\hat\sigma_z\frac{\hbar^2k_\perp}{m}\frac{\partial}{\partial s}
  +\hat\sigma_x G_R^{(\alpha)}-\hat\sigma_y G_I^{(\alpha)}
  +\varepsilon_d\right]\hat w=\varepsilon\hat w\ ,
\end{equation}
Here
\begin{equation}
\label{Doppler}
  \varepsilon_d=-\frac{\hbar^2k_\perp}{m}\frac{\alpha b}{s^2+b^2}
\end{equation}
is the Doppler shift, and
\begin{eqnarray}
\label{g-Def}
  \nonumber
  G_R^{(\alpha)}=D_M\left(\sqrt{s^2+b^2}\right){\rm Re}
  \left\{\left[\frac{s+ib}{\sqrt{s^2+b^2}}\right]^{M-2\alpha}\right\},\\
  G_I^{(\alpha)}=D_M\left(\sqrt{s^2+b^2}\right){\rm Im}
  \left\{\left[\frac{s+ib}{\sqrt{s^2+b^2}}\right]^{M-2\alpha}\right\}\,,
\end{eqnarray}
are the real and imaginary part of the off--diagonal potential,
correspondingly. One can see that choosing  $M-2\alpha$ to be an odd
positive integer we can change the parity of the potentials in the
Hamiltonian so that $G_R^{(\alpha)}(s)$ is an odd function. Comparing Eq.
(\ref{Quasiclass-Doppler}) and Eq. (\ref{Quasiclass}) we observe that the
above gauge transformation also produces a Doppler shift of the energy
levels. In principle, both the Doppler shift term and the term
$\hat\sigma_yG_I^{(\alpha)}$ are not small and can be of the order of
$\Delta_0$ if the impact parameter is rather large: $b\sim\xi$. Thus,
strictly speaking we can consider the expression (\ref{sol0}) with $G_R$
replaced by $G_R^{(\alpha)}(s)$ as a zero--order approximation and use the
perturbation technique discussed above only provided the energy
corrections arising from the terms $\hat\sigma_yG_I^{(\alpha)}$ and
$\varepsilon_d$ almost compensate each other. The anomalous branches which
we study cross the Fermi level at certain impact parameters
$-\mu_j/k_\perp$, thus, the perturbation method should be adequate in the
vicinity of these points. Changing $\alpha$ in the interval
$0\le\alpha<M/2$ we get a set of possible odd $M-2\alpha$ values providing
us a set of different zero--order approximations which allow to obtain the
spectrum as a function of $b$. Surprisingly, we shall see below that this
method can describe the spectrum behavior even beyond its validity domain,
i.e. when the energy is comparable to $\pm\Delta_0$.

It is convenient to parameterize the wave functions as
\begin{equation}
\label{zeta-eta-Def}
  \hat w =e^{\zeta(s)}\left(
  \begin{array}{c}
    e^{i\eta(s)/2} \\
    e^{-i\eta(s)/2} \\
  \end{array}\right)\ .
\end{equation}
As a result, we have the equations
\begin{eqnarray}
\label{eta-Eq}
  \nonumber
  \frac{\hbar^2k_\perp}{2m}\frac{\partial\eta}{\partial s}
  &+&G_R^{(\alpha)}\cos\eta+G_I^{(\alpha)}\sin\eta
  = \varepsilon-\varepsilon_d\,,\\
\label{zeta-Eq}
  \frac{\hbar^2k_\perp}{m}\frac{\partial\zeta}{\partial s}
  &+&G_R^{(\alpha)}\sin\eta-G_I^{(\alpha)}\cos\eta=0\,.
\end{eqnarray}
Considering the localized states one should take the following boundary
conditions for odd $M-2\alpha$ values:
\begin{equation}
\label{eta-bound-Cond}
  \cos\eta(\pm\infty)=\pm\varepsilon/\Delta_0\,,\ \
  \sin\eta(\pm\infty)=\sqrt{1-\varepsilon^2/\Delta_0^2}\,.
\end{equation}
In order to construct the solution we note that the mutual phase $\eta$ of
the electron and hole components in the zero-order solution (\ref{sol0})
is constant [$\eta(s)=\pi/2$], therefore, within the perturbation theory
we can linearize the Eq. (\ref{eta-Eq}) for $\eta$ close to $\pi/2$
introducing $\tilde\eta=\pi/2-\eta$:
\begin{equation}
\label{chi-Eq}
  \frac{\partial\tilde\eta}{\partial s}
  -\frac{2m}{\hbar^2k_\perp}G_R^{(\alpha)}\tilde\eta
  =\frac{2m}{\hbar^2k_\perp}
  \left(\varepsilon_d+G_I^{(\alpha)}-\varepsilon\right)\,.
\end{equation}
To exclude the divergent solutions of this equation we should impose the
integral condition describing the anomalous spectral branches:
\begin{equation}
\label{Spectrum-alpha}
  \varepsilon_M^{(\alpha)}=\frac{\int_0^\infty
  \left[\varepsilon_d+G_I^{(\alpha)}(s)\right]e^{-K(s)}\,ds}
  {\int_0^\infty e^{-K(s)}\,ds}\ ,
\end{equation}
where
\begin{equation}
\label{K-def}
  K(s)=\frac{2m}{\hbar^2k_\perp}\int_0^s G_R^{(\alpha)}(t)\,dt \,.
\end{equation}

\begin{figure}[htb]
\centerline{\includegraphics[width=1.0\linewidth]{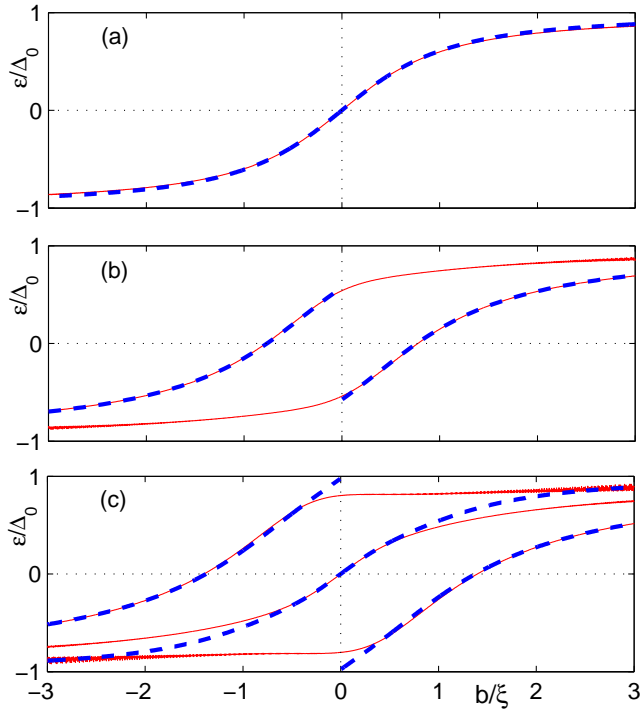}}
 \caption{\label{Fig:Spectrum-M=2} (Color online).
The anomalous spectral branches as functions of the impact parameter $b$
for $k_z=0$, obtained from Eq. (\ref{Spectrum-alpha}) (blue dash lines)
for $M=1$ (a), $M=2$ (b), $M=3$ (c). The anomalous spectral branches
obtained from numerical solution of the eigen--value problem (\ref{BdG})
are shown by solid red lines. The gap profiles are approximated as
$D_M(r)=\Delta_0(r/\sqrt{r^2+\xi^2})^M$ with the parameter $k_F\xi=200$.}
\end{figure}

Taking ${D_1(r)=\Delta_0 r/\sqrt{r^2+\xi^2_v}}$ and $\alpha=0$ for the
simplest case of a singly--quantized vortex ($M=1$) we get an explicit
expression for the CdGM spectrum:
\begin{equation}
\label{Spectrum:M=1}
  \varepsilon_1^{(0)}=\frac{\Delta_0b}{\sqrt{b^2+\xi_v^2}}
  \frac{\mathcal{K}_0
  \left(2m\Delta_0\sqrt{b^2+\xi_v^2}/(\hbar^2k_\perp)\right)}
  {\mathcal{K}_1
  \left(2m\Delta_0\sqrt{b^2+\xi_v^2}/(\hbar^2k_\perp)\right)}\,,
\end{equation}
where $\mathcal{K}_n$ is the McDonald function. To obtain the spectrum for
$M=2$ we need to consider the case $\alpha=1/2$ so that to reduce the
problem to the one for a unity vorticity and a certain Doppler shift. For
the case $M=3$ the set of anomalous branches can be obtained taking two
$\alpha$ values: $\alpha=0$ and $\alpha=1$. The solution
$\varepsilon_3^{(0)} (b)$ gives us the branch crossing the Fermi level at
zero impact parameter, while  two other branches are described by
$\varepsilon_3^{(1)} (b)$. The typical plots of quasiparticle spectra for
vortices with winding numbers $M=1,2,3$  are shown in Fig.
\ref{Fig:Spectrum-M=2}.
 %
 %
Comparing the spectra $\varepsilon_M^{(\alpha)}$ with the branches
obtained from our direct numerical analysis of BdG equations (\ref{BdG})
one can see that our perturbation method provides a reasonable description
of the low energy spectrum behavior. The numerical solution of the
eigen--value problem (\ref{BdG}) was carried out using a representation of
the BdG operator in the truncated basis of the normal--metal
eigenfunctions. We solve the system of linear equations which correspond
to the boundary conditions at the superconductor/insulator boundary of a
cylinder with a finite radius (we took $R=7\xi$). The small oscillations
of the spectrum as a function of $b$ (solid red lines in the Fig.
\ref{Fig:Spectrum-M=2}) result from the interference of incident and
reflected from the boundary quasiparticle waves~\cite{PRL-2005}. Note,
that our perturbation procedure fails to describe proper behavior of all
the branches in the vicinity of the gap value $\Delta_0$: e.g., for $M=2$
the function $\varepsilon_2^{(1/2)}$ jumps at $b=0$ from the upper branch
to the lower one and, thus, we can not describe the part of the upper
branch approaching $\Delta_0$ for $b>0$.

To obtain the spectrum as a function of discrete variable $\mu$ instead of
a continuous $b$ we should apply the Bohr--Sommerfeld quantization rule
for the angular momentum [see Eq. (\ref{bohr})] with $\beta=\{M/2\}$ which
results from the obvious condition that the wave function $\hat g$ is
single--valued. Here $\{...\}$ denotes the fractional part. For the odd
(even) vorticity $M$ we obtain $\mu=n+1/2$ ($\mu=n$), where $n$ is an
integer.

\section{Decay of a multiquantum vortex into a set of separated
vortices}
 \label{SEC:molecule}

In this section we consider modification of the quasiparticle spectrum
caused by the decay of a multiquantum vortex into a set of separated
singly--quantized vortices, which occurs under the magnetic field
decreasing. For simplicity sake we restrict ourselves to the case of
two--vortex system with a certain intervortex distance $a$ controlled by
the external magnetic field. The case $a=0$ corresponds to a
doubly--quantized vortex while the limit $a\gg\xi$ corresponds to a pair
of isolated singly--quantized vortices. In this section we again neglect
the effect of normal quasiparticle scattering at the sample boundary.

\subsection{Quasiclassical consideration}
As a first step of our analysis we choose to apply the approximate
quasiclassical procedure developed in the previous section. It is natural
to expect that the validity range for this method should be restricted to
the region of rather small distances $a<a_c$ when one can neglect the
Landau--Zener tunneling between the quasiclassical orbits $\mu(\theta_p)$
described in Section \ref{SEC:picture}. To describe the system of two
singly--quantized vortices positioned at ${\bf r}=\pm{\bf a}/2=\pm(a/2,0)$
we  fit the gap function as follows:
\begin{eqnarray}
\label{2VortMolecGap}
 \nonumber
 \Delta({\bf r})
 =\Delta_0f_1({\bf r}-{\bf a}/2)f_1({\bf r}+{\bf a}/2)\\
 =\Delta_0
 \left|f_1({\bf r}-{\bf a}/2)\right|\left|f_1({\bf r}+{\bf a}/2)\right|\\
 \nonumber
 \times\frac{x+iy-a/2}{|x+iy-a/2|}\frac{x+iy+a/2}{|x+iy+a/2|}\,,
\end{eqnarray}
where $f_1({\bf r})$ is a normalized gap function of a singly--quantized
vortex. It is convenient to rewrite the above expression as a
superposition of functions with two different vorticities:
\begin{equation}
\label{Clem-molec}
  \Delta({\bf r})=\Delta_0
  \left[f_2({\bf r})e^{2i\theta}+f_0({\bf r})\right]\ .
\end{equation}
Taking the simplest core model
\begin{equation}
\label{Clem-profile}
  \left|f_1({\bf r})\right|=\frac{r}{\sqrt{r^2+\xi_v^2}}
\end{equation}
with the core size $\xi_v$ we obtain:
\begin{eqnarray}
  \nonumber
  f_2({\bf
  r})=\frac{x^2+y^2}{\sqrt{(x^2+y^2+\xi_v^2+a^2/4)^2-a^2x^2}}\,,\\
  \nonumber
  f_0({\bf r})=-\frac{a^2/4}{\sqrt{(x^2+y^2+\xi_v^2+a^2/4)^2-a^2x^2}}\ .
\end{eqnarray}

To solve the equations (\ref{andreev-eqs}) with the gap function given by
(\ref{Clem-molec}) we apply the gauge transformation
(\ref{Gauge-transform}) with $\alpha=1/2$. The expression for the
quasiclassical spectrum takes the form (\ref{Spectrum-alpha}) with
\begin{eqnarray}
\label{gRgI-molec}
  G_R&=&\Delta_0\left[f_2(x,y)\frac{s}{\sqrt{s^2+b^2}} \right.\\
  \nonumber
  &+&f_0(x,y)\left.
  \frac{s\cos(2\theta_p)-b\sin(2\theta_p)}{\sqrt{s^2+b^2}}\right]\,,\\
  G_I&=&\Delta_0\left[f_2(x,y)\frac{b}{\sqrt{s^2+b^2}} \right.\\
  \nonumber
  &+&f_0(x,y)\left.
  \frac{s\sin(2\theta_p)+b\cos(2\theta_p)}{\sqrt{s^2+b^2}}\right]\,,
\end{eqnarray}
where $(x,y)$ variables are given by Eqs. (\ref{Coord}).

\begin{figure}[htb]
\centerline{\includegraphics[width=1.0\linewidth]{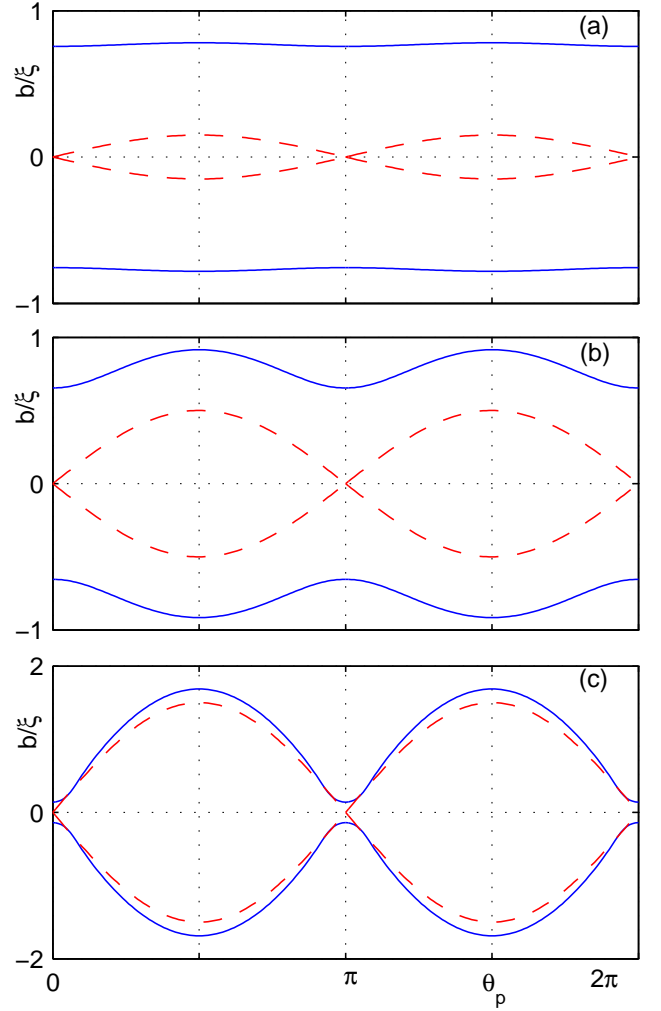}}
  \caption{\label{Fig:b-a-molec}
The isoenergetic curves $b(\theta_p)$ for a two--vortex system. We choose
here $\varepsilon=0$, $k_z=0$, $a=0.3\xi$ (a), $a=\xi$ (b), $a=3\xi$ (c).
The dash curves correspond to the isoenergetic lines for two
non--interacting singly--quantized vortices. The gap is approximated by
Eqs. (\ref{2VortMolecGap}),~(\ref{Clem-profile}) with $\xi_v=\xi$.}
\end{figure}

\begin{figure}[htb]
\centerline{\includegraphics[width=1.0\linewidth]{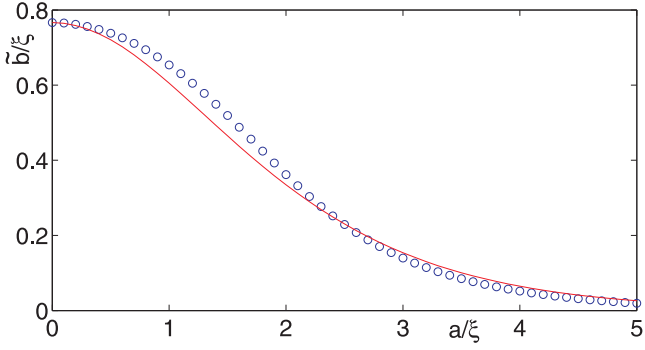}}
 \caption{\label{Fig:b-tilde}
Splitting of quasiclassical orbits $\tilde{b}$ obtained from Eq.
(\ref{K0-Clem}) (solid line) and found from the spectrum of a two--vortex
system (\ref{Spectrum-alpha}) (circles) for $k_z=0$. The gap profile is
approximated by Eq. (\ref{Clem-molec}) with $\xi_v=\xi$.}
\end{figure}

The resulting dependencies of the impact parameter vs $\theta_p$ for zero
energy and different intervortex distances are shown in Fig.
\ref{Fig:b-a-molec}.
 %
 %
These calculations of the low energy part of the spectrum appear to be in
a good agreement with the two--level model (\ref{orbit1}) and, thus, the
angular dependence of the impact parameter can be fitted by the
expression:
\begin{equation}
\label{b-approx}
  b(\theta_p)=\frac{\varepsilon}{\tilde{\omega}(k_z,a)k_\perp}
  \pm\sqrt{\tilde{b}^2(k_z,a)+\left(\frac{a}{2}\sin\theta_p\right)^2}\,,
\end{equation}
where $\tilde{\omega}(k_z,a)\sim\Delta_0/k_\perp\xi$, and
$\tilde{b}=\delta\varepsilon/(k_\perp\omega)$ is the splitting of
quasiclassical orbits. In the limit $a=0$ we get the spectrum of a
doubly--quantized vortex: $\varepsilon= \tilde{\omega}(k_z,0)[b\pm
\tilde{b}(k_z,0)]$, where $\tilde{b}(k_z,0)$  is of the order of $\xi$ for
small $k_z$ values. For large intervortex distances the value $\tilde{b}$
is exponentially small: $\tilde{b}\sim\xi\exp(-k_F a/(k_\perp\xi))$ [see
Eq. (\ref{delta-mu})]. Generally, in the whole interval of distances $a$
the splitting of quasiclassical orbits is defined by the overlapping of
the wave functions localized in the cores of neighboring singly--quantized
vortices. This overlapping is determined by the factor $\exp(-K_0(s))$
describing the decay of the wave function in a singly--quantized vortex
(see Ref. \onlinecite{CdGM}):
\begin{equation}
\label{DL-defin}
  K_0(s)=\frac{2m}{\hbar^2k_\perp}\int_0^s\Delta(t)\,dt\,.
\end{equation}
Thus, the spectrum (\ref{Spectrum-alpha}) can be fitted by the one
describing a two--level system if we put $\tilde{b}=\tilde
b(k_z,0)\exp(-K_0(a/2))$. Taking the vortex core model
(\ref{Clem-profile}) we obtain:
\begin{equation}
\label{K0-Clem}
 \tilde{b}=\tilde b(k_z,0)\exp\left(-2\frac{k_F}{k_\perp}\frac{\xi_v}{\xi}
  \left[\sqrt{\frac{a^2}{4\xi_v^2}+1}-1\right]\right)\,.
\end{equation}
Comparing this approximate expression with the orbit splitting calculated
using the spectrum (\ref{Spectrum-alpha}) (see Fig. \ref{Fig:b-tilde} for
the particular case $k_z=0$) we can find appropriate $\tilde b(k_z,0)$
values.
 %

To find the quantized energy levels we can use the Bohr--Sommerfeld
quantization rule (\ref{bohr}) which takes the form $
S(\varepsilon,k_z)=2\pi(n+\beta)$, where
\begin{equation}
  S(\varepsilon,k_z)= \int_0^{2\pi}\mu(\theta_p)\,d\theta_p
  =-k_\perp\int_0^{2\pi}b(\theta_p)\,d\theta_p
\end{equation}
is the area under the isoenergetic line $\mu(\varepsilon,k_z)$ in the
$(\mu,\theta_p)$--plane. Calculating this integral for the two--level
model (\ref{b-approx}) we obtain:
\begin{equation}
\label{S-elliptic}
  S(\varepsilon,k_z)=
  -2\pi\frac{\varepsilon}{\tilde{\omega}}
  \pm 2k_\perp\sqrt{a^2+4\tilde{b}^2}\,
  {\rm E}\left(\frac{a}{\sqrt{a^2+4\tilde{b}^2}}\right),
\end{equation}
where ${\rm E}(k)=\int_0^{\pi/2}\sqrt{1-k^2\sin^2\theta}\,d\theta$ is the
complete elliptic integral of the second type. As a result, we find the
quasiparticle spectrum of a two--vortex system
\begin{equation}
\label{Spectrum-molec-n}
  \varepsilon_n = \tilde{\omega}\left[-n-\beta\pm
  \frac{k_\perp\sqrt{a^2+4\tilde{b}^2}}{\pi}
  {\rm E}\left(\frac{a}{\sqrt{a^2+4\tilde{b}^2}}\right)\right].
\end{equation}
The spectrum (\ref{Spectrum-molec-n}) is analogous to the spectrum of the
doubly--quantized vortex (\ref{Volovik-spectr}) and is drastically
different from the CdGM spectrum. The case $a=0$ gives us the  spectrum
(\ref{Volovik-spectr}) of a doubly--quantized vortex. Treating the
opposite limit $a\gg\tilde{b}$ with the logarithmic accuracy we obtain:
\begin{equation}
\label{S-well-separated}
  S(\varepsilon,k_z)=-2\pi\frac{\varepsilon}{\tilde{\omega}}\pm 2k_\perp a
  \left[1+2\left(\frac{\tilde{b}}{a}\right)^2\ln\left(\frac{a}{\tilde{b}}
  \right)\right]\,,
\end{equation}
and
\begin{equation}
\label{Spectrum-well-separated}
  \varepsilon_n = \tilde{\omega}\left\{-n\pm \frac{k_\perp a}{\pi}\left[
  1+2\left(\frac{\tilde{b}}{a}\right)^2\ln\left(\frac{a}{\tilde{b}}\right)
  \right]\right\}\,.
\end{equation}

\subsection{Landau-Zehner tunneling between quasiclassical orbits.}
 \label{SEC:moleculeSub:beyond}

The above description is valid provided the intervortex distance is rather
small ($a<a_c$) when the splitting $\delta\mu$ between isoenergetic lines
$\mu(\theta_p)$ is large compared to the quantum mechanical uncertainty of
the angular momentum $\Delta\mu$ and the probability of Landau--Zehner
tunneling between quasiclassical orbits can be neglected [see
Eq.(\ref{LZest1})]. As a next step, we proceed with a quantitative
analysis of quasiparticle spectrum of the two--vortex system in case of
the small splitting of the isoenergetic lines in $(\mu,\theta_p)$--plane,
when the vortices are well separated so that $a\gtrsim a_c$. In order to
calculate the quasiparticle spectrum taking into account the influence of
Landau--Zehner tunneling we should go beyond the quasiclassical
consideration of the angular precession of quasiparticle trajectories. It
means that we can not neglect the non--commutativity of canonical
variables: $[\hat\mu,\theta_p]=-i$, where $\theta_p$ is the trajectory
orientation angle and $\hat\mu=-i\partial/\partial\theta_p$ is the angular
momentum operator. Keeping in mind the symmetry of the gap function in a
two--vortex system we can reduce the problem to the one describing a
single vortex with an additional boundary condition imposed on the wave
function at the plane $x=0$ positioned between vortices. Indeed, the gap
function distribution corresponding to the two--vortex system possesses
the following symmetry: $\Delta(x,y)=\Delta^*(-x,y)$. As a result, for the
eigenfunctions we obtain:
\begin{equation}\label{symm}
  \hat\Psi(x,y)=e^{i\chi}\hat\Psi^*(-x,y),
\end{equation}
where $\chi$ is a constant phase.  The spectrum does not depend on $\chi$,
since for any eigenfunction $\hat\Psi$ satisfying Eq. (\ref{symm}) we can
introduce a new function $\hat\Psi_1=\hat\Psi e^{-i\chi/2}$, which
corresponds to the same energy level and has the following symmetry:
$\hat\Psi_1(x,y)=\hat\Psi_1^*(-x,y)$. Therefore we can choose $\chi=0$ and
obtain the boundary conditions at the plane $x=0$:
\begin{equation}
\label{bcvv}
  \hat\Psi=\hat\Psi^*;\qquad
  \frac{\partial\hat\Psi}{\partial x}
  =-\frac{\partial\hat\Psi^*}{\partial x}\,.
\end{equation}
For the sake of simplicity we neglect the anisotropy of the gap function
around the vortex positioned in the half--space $x<0$. Nevertheless the
solution can not be characterized by a definite angular momentum because
of the above boundary condition responsible for interaction of different
angular harmonics. Thus, following Ref. \onlinecite{PRB-2007} we introduce
the angular--momentum expansion for the solution:
\begin{equation}
\label{AngMom}
  \hat\psi( s,\theta_p)=\sum_\mu e^{i\mu\theta_p+i\hat\sigma_z
  \theta_p/2}\hat g_{\mu} (s),
\end{equation}
where $\mu=n+1/2$, and $n$ is an integer. The function $\hat g_{\mu}$
satisfies the Andreev equation (\ref{andreev-eqs}) along the
quasiclassical trajectory with $b=-\mu/k_\perp$. For small impact
parameters $b\ll\xi$ the Eq. (\ref{andreev-eqs}) can be solved
analytically, yielding a general expression for the function $\hat
g_\mu(s)$ in the following form:
\begin{eqnarray}
  \nonumber
  \hat g_\mu(s)=c_\mu\hat G_{1\mu}(s)+d_\mu\hat G_{2\mu}(s),
\end{eqnarray}
 where $c_\mu$, $d_\mu$ are arbitrary constants and we choose the
 fundamental solutions so that $G_{1\mu}(-\infty)=0$ while
 $G_{2\mu}(+\infty)=0$ (see Ref. \onlinecite{PRB-2007}):
\begin{eqnarray}
  \nonumber
  \hat G_{1\mu}=\left[e^{-|K_0(s)|/2}-i\frac{\gamma}{2}
  ({\rm sgn}s+1)\hat\sigma_ze^{|K_0(s)|/2}\right]\hat\lambda\,,\\
  \nonumber
  \hat G_{2\mu}=\left[e^{-|K_0(s)|/2}-i\frac{\gamma}{2}
  ({\rm sgn} s-1)\hat\sigma_ze^{|K_0(s)|/2}\right]\hat\lambda\,.
\end{eqnarray}
Here $\hat\lambda=(\exp(i\pi/4),\exp(-i\pi/4))$,
\begin{equation}
  \gamma(\mu)=\frac{\Lambda}{\Delta_0}
  \left[\varepsilon(\mu)-\varepsilon\right]\,,
\end{equation}
\begin{equation}
  \label{lamb}
  \Lambda=\frac{2k_F}{k_\perp\xi}\int_0^\infty e^{-K_0(s)}\,ds\,.
\end{equation}
Here the CdGM spectrum is taken as a linear function of $|\mu|$ for small
$\mu\ll k_\perp\xi_v$: $\varepsilon (\mu)=-\mu\omega$ with interlevel
spacing
\begin{eqnarray}
  \nonumber
  \omega=\frac{1}{\Lambda}\frac{2k_F}{k^2_\perp\xi}\int_0^{\infty}
  \frac{\Delta(s)}{s}e^{-K_0(s)}ds\,.
\end{eqnarray}
It is convenient to introduce the angle--dependent functions:
\begin{eqnarray}
  \nonumber
  C(\theta_p)=\sum_\mu e^{i\mu\theta_p}c_{\mu}\,,\qquad
  D(\theta_p)=\sum_\mu e^{i\mu(\theta_p+\pi)}d_{\mu}\,.
\end{eqnarray}
These functions appear to be nonzero only in the angular interval
$-\pi/2<\theta_p<\pi/2$ because of the decay of the wave function $\hat
\Psi(x,y)$ in the left half-space far from the vortex. At the boundaries
of this angular interval we should impose the conditions
\begin{equation}
\label{bcVV}
  C(\pm\pi/2)=\pm D(\mp\pi/2)\,,
\end{equation}
which result from the continuity of the wave function $\psi(s,\theta_p)$.

Within the large angle domain $|\theta_p|\gg\xi/a$ the wave function
$\hat\psi(s,\theta_p)$ can be found using a tight binding approximation as
a sum of two single--vortex solutions localized on vortices at ${\bf
r}=\pm (a/2,0)$ (see Ref. \onlinecite{Melnikov-Silaev-2006}). Comparing
the tight--binding solution with Eq.(\ref{AngMom}) we obtain that
$\sum\gamma(\mu)e^{i\theta_p\mu}c_\mu=0$ and
$\sum\gamma(\mu)e^{i(\theta_p+\pi)\mu}d_\mu=0$, or
\begin{equation}
  \label{CD}
  C(\theta_p),D(\theta_p-\pi)\sim e^{-i\varepsilon\theta_p/\omega}.
\end{equation}

The deviations of angular functions $C(\theta_p),D(\theta_p)$ from
(\ref{CD}) due to the intervortex quasiparticle tunneling  occur in the
narrow angular domain $|\theta_p|\ll\xi/a$. Within this domain we apply
the stationary phase method to handle the integral in Eq.(\ref{xy}). For a
given value of angular momentum $\mu$ the stationary phase points are
given by $\sin(\theta_p-\theta)=\mu/[k_\perp R(\theta)]$, where
$R(\theta)= a/(2\cos\theta)$ corresponds to the line $x=0$ in the polar
coordinate system with the origin in the vortex center. Assuming $|\mu|\ll
k_\perp a$ and $|\theta|\ll \pi$, we obtain the stationary phase points:
$\theta_{p1}\approx\theta+2\mu/(k_\perp a)$ and
$\theta_{p2}=\pi+\theta-2\mu/(k_\perp a)$. Then, the expression for the
wave function $\Psi (r,\theta)$ at $r=R(\theta)$ and $|\theta|\ll\pi$
reads as follows:
\begin{eqnarray}
  \nonumber
  \hat\Psi(R(\theta),\theta)
  =e^{i\varphi}\int_{-\infty}^{\infty} e^{i\mu^2/k_\perp a}
  \left(e^{-K_0(a/2)}-i \gamma\hat\sigma_z\right)
  \hat\lambda e^{i\mu\theta}c_\mu d\mu\\
  \nonumber
  - \hat\sigma_z e^{-i\varphi}\int_{-\infty}^{\infty} e^{-i\mu^2/k_\perp a}
  \left(e^{-K_0(a/2)}+i\gamma\hat\sigma_z \right)
  \hat\lambda e^{i\mu(\theta+\pi)}d_\mu d\mu\,,
\end{eqnarray}
where $\varphi=k_\perp a(1+\theta^2/2)/2$ and the discreteness of angular
momentum $\mu$ is neglected. Then, from Eqs. (\ref{bcvv}) we obtain:
\begin{equation}
  \label{r1}
  {\rm Im}\left[e^{i\varphi}\int_{-\infty}^\infty e^{i\mu^2/k_\perp a}
  \left(e^{-K_0(a/2)}-i\gamma\hat\sigma_z\right)
  \hat\lambda e^{i\mu\theta}c_\mu d\mu\right]=0\,,
  \end{equation}
which yields the following equation (see Appendix \ref{ScaleTr}):
\begin{eqnarray}
  \label{r11}
  \nonumber
  \left(e^{-K_0(a/2)}-i\hat\sigma_z\hat\gamma\right)
  \hat\lambda C_2(\theta)\\
  =e^{-2i\tilde{\varphi}}\left(e^{-K_0(a/2)}+i\hat\sigma_z\hat\gamma\right)
  \hat\lambda^*C_1(\theta)\,,
\end{eqnarray}
where $\tilde{\varphi}=k_\perp a(1-\theta^2/2)/2$,
$C_2(\theta)=iC(\theta)$ and $C_1(\theta)=C^*(-\theta)$. Then, Eq.
(\ref{r11}) can be written as follows:
\begin{eqnarray}
\label{tunneling1}
  \nonumber
  \left(i\omega\frac{\partial}{\partial\theta}-\varepsilon\right)
  C_2(\theta)&=&e^{-2i\tilde{\varphi}}J C_1(\theta)\,,\\
  \left(i\omega\frac{\partial}{\partial\theta}-\varepsilon\right)
  C_1(\theta)&=&e^{2i\tilde{\varphi}}J C_2(\theta)\,,
\end{eqnarray}
where the overlap integral is
\begin{equation}
\label{J-def}
  J=\frac{\Delta_0}{\Lambda}e^{-K_0(a/2)}\,.
\end{equation}
Analogously, for the functions $D_2(\theta)=iD(\theta)$ and
$D_1(\theta)=D^*(-\theta)$ we obtain:
\begin{eqnarray}
\label{tunneling2}
  \nonumber
  \left(i\omega\frac{\partial}{\partial\theta}-\varepsilon\right)
  D_2(\theta)&=&e^{2i\tilde{\varphi}}J D_1(\theta)\,,\\
  \left(i\omega\frac{\partial}{\partial\theta}-\varepsilon\right)
  D_1(\theta)&=&e^{-2i\tilde{\varphi}}J D_2(\theta)\,.
\end{eqnarray}

\begin{figure}[htb]
\centerline{\includegraphics[width=1.0\linewidth]{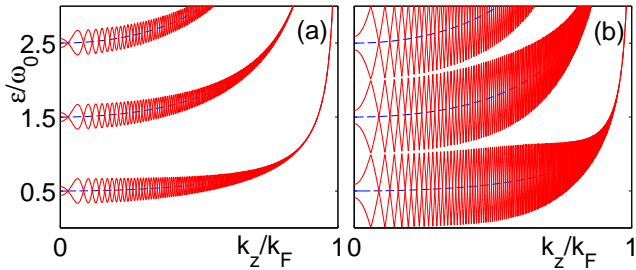}}
 \caption{\label{FigSpectrVV}
The quasiparticle spectrum of two vortices calculated using the Eq.
(\ref{spectrumVV}) for $a=5\xi$ (a) and $a=3.5\xi$ (b). The CdGM spectrum
is shown by the dash lines. The vortex core profile for a single vortex is
approximated by Eq. (\ref{Clem-profile}) with $\xi_v=\xi$, $k_F\xi=200$.}
\end{figure}

Considering the Eqs. (\ref{tunneling1}), (\ref{tunneling2}) in the angular
domain $(\xi/a) \gg|\theta|\gg (\xi/a)J/\Delta_0$ we can neglect the
rapidly oscillating right hand side, therefore the asymptotic form of
angular functions ${\bf C}$ and ${\bf D}$ corresponds to the
non--interacting vortices:
\begin{eqnarray}
\label{iVort}
  \nonumber
  {\bf C}&=&(C_1,C_2)
  =e^{-i\varepsilon\theta/\omega}(\tilde{c}_1,\tilde{c}_2)\,,\\
  {\bf D}&=&(D_1,D_2)
  =e^{-i\varepsilon\theta/\omega}(\tilde{d}_1,\tilde{d}_2)\,,
\end{eqnarray}
where $\tilde{c}_1$, $\tilde{c}_2$, $\tilde{d}_1$, $\tilde{d}_2$ are
arbitrary constants. Then, solving Eqs. (\ref{tunneling1}),
(\ref{tunneling2}) we can find the transfer matrix $\hat X$ matching the
large--angle asymptotics:
\begin{eqnarray}
  \nonumber
  {\bf C}(\theta_{th})=\hat X {\bf C}(-\theta_{th})\,,
\end{eqnarray}
where $\hat X$ is a transfer matrix, and $\theta_{th}\sim\xi/a$ is a
threshold angle where we match the solutions for the large--angle and
small--angle domains. Introducing new functions
\begin{eqnarray}
  \nonumber
  B_2(\theta)=C_2(\theta)e^{i\tilde{\varphi}-i\varepsilon\theta/\omega}\,,
  \quad
  B_1(\theta)=C_1(\theta)e^{-i\tilde{\varphi}-i\varepsilon\theta/\omega}\,,
\end{eqnarray}
we obtain the equations:
\begin{eqnarray}
\label{TM1}
  \nonumber
  \left(i\frac{\partial}{\partial\tilde{\theta}}+\tilde{\theta}\right)B_1
  =pB_2\,,\\
  \left(i\frac{\partial}{\partial\tilde{\theta}}-\tilde{\theta}\right)B_2
  =pB_1\,,
\end{eqnarray}
where
\begin{equation}
\label{p-def}
  p=\frac{J}{\omega\sqrt{k_\perp a/2}}
\end{equation}
and $\tilde{\theta}=\theta\sqrt{k_\perp a/2}$. This equations coincide
with equations obtained in Ref. \onlinecite{Melnikov-Silaev-2006}. The
problem described by the Eqs. (\ref{TM1}) is equivalent to the one
describing the interband tunneling~\cite{kane} or one--dimensional motion
of a Dirac particle in a uniform electric field and the solution can be
written in terms of the parabolic cylinder functions (see Appendix
\ref{TransferMat}), yielding the transfer matrix:
\begin{equation}
  \label{TransM}
  \hat X=e^{-\pi p^2/2}\hat I+i\left(\hat\sigma_y
  \rm{Re}\tau+\hat\sigma_x\rm{Im}\tau \right)\,.
\end{equation}
Here $\hat I$ is the unity matrix,
\begin{eqnarray}
  \nonumber
  \tau&=&\sqrt{2\sinh(\pi p^2/2)}e^{-\pi p^2/4}e^{i\chi}\,,\\
  \nonumber
  \label{chi-spectrum}
  \chi&=&k_\perp a+p^2\ln\left(\theta_{th}\sqrt{k_\perp a}\right)\\
  &+&\arg\left(\Gamma\left(1-i\frac{p^2}{2}\right)\right)+\frac{\pi}{4}\,,
\end{eqnarray}
and $\Gamma$ is the gamma function. Analogously, for the functions ${\bf
D}(\theta)$ we find:
\begin{eqnarray}
  \nonumber
  {\bf D}(\theta_{th})=\hat\sigma_x\hat X\hat\sigma_x{\bf D}(-\theta_{th})\,.
\end{eqnarray}
Matching the wave function in different angular domains and using the
boundary conditions (\ref{bcVV}) we obtain the spectrum:
\begin{equation}
\label{spectrumVV}
  \cos(\pi\varepsilon/\omega)=\pm e^{-\pi p^2/4}
  \sqrt{2\sinh(\pi p^2/2)}\sin\chi\,.
\end{equation}
Within the present theory we can not determine the threshold angle
$\theta_{th}$ precisely. However, since the dependence of $\chi$ on
$\theta_{th}$ is logarithmic, the $\theta_{th}$--dependent term in Eq.
(\ref{chi-spectrum}) can be considered as an additional constant phase of
the oscillations of the energy levels. The spectrum of two vortices
calculated using the Eq. (\ref{spectrumVV}) for small ($a>a_c$) and large
($a<a_c$) intervortex distance is shown in Fig. \ref{FigSpectrVV}.
 %
 %
One can see that the transformation of the spectrum $\varepsilon(k_z)$
occurs according to the scenario suggested above: as we decrease the
distance $a$ below $a_c$ the crossover to the doubly--quantized vortex
spectrum starts in the region of small $k_z$ values defined by the
condition $p>1$. For $p\ll1$ we get the CdGM spectrum with a small
oscillatory correction:
\begin{equation}
\label{SpectrumSmall}
  \varepsilon\simeq\omega[n+1/2\mp(-1)^n(p/\sqrt{\pi})
  \sin(k_\perp a +\pi/4)]\,.
\end{equation}
The minigap $\varepsilon_{min}=[1-2p/\sqrt{\pi}]\omega_0/2$ vanishes for
$a=a_c$.

For the large values $p\gg1$ the Eq. (\ref{spectrumVV}) yields the
spectrum in the form:
\begin{equation}
\label{SpectrumLarge}
  \varepsilon=\omega\left[n\pm\frac{k_\perp a
  +p^2\ln\left(\theta_{th}\sqrt{k_\perp a/p^2}\right)}{\pi}\right]\,.
 \end{equation}
Here we have used the asymptotic formula for the argument of gamma
function at $p\gg 1$:
$\arg(\Gamma(1-ip^2/2))\approx-(p^2/2)[\ln(p^2/2)-1]$. This result can be
formulated as the Bohr--Sommerfeld quantization rule
$S(\varepsilon,k_z)=2\pi n$ with
\begin{equation}
\label{areaVV}
  S(\varepsilon,k_z)=2\pi\frac{\varepsilon}{\omega}\pm
  \left[2k_\perp a
  +2p^2\ln\left(\theta_{th}\sqrt{\frac{k_\perp
  a}{|p|^2}}\right)\right]\,.
\end{equation}
This result coincides with the formula (\ref{S-well-separated}) obtained
by the evaluation of the Bohr--Sommerfeld integral at quasiclassical
orbits $\mu (\theta_p)$ such as ones shown in Fig. \ref{fig1}(b).

\section{Boundary effects: vortex near the surface}
\label{SEC:boundary}

Now we proceed with a quantitative analysis of the effect of normal
reflection of quasiparticles at the boundary. We consider a vortex near a
smooth surface approximated by a parabolic cylinder (see Sec.
\ref{reflection}). Considering the quasiclassical trajectories which are
rather far from being parallel to the system optical axis we find that
either incident or reflected trajectory appears to pass far from the
vortex core. The quasiclassical spectrum for this case is the same as for
a single vortex:
\begin{equation}
\label{QSpectrIv}
  \varepsilon(b,\theta_p,k_z)=-\omega k_\perp(b-d\sin \theta_p)\,.
\end{equation}
This spectrum should be strongly disturbed for trajectories passing close
to the optical axis when both the incident and reflected trajectories pass
through the vortex core. In this case the trajectory orientation angles
should be taken in the domains $|\theta_p|<\xi/d$ or
$|\pi-\theta_p|<\xi/d$ and the impact parameters defined relative to the
point with coordinates $r=d$ and $\theta=0$ (see Sec.\ref{reflection} for
notations) should be rather small: $|b|\ll d$. Solving Eq.
(\ref{andreev-eqs}) along the incident and reflected trajectories and
matching the solutions to meet the zero boundary condition for the wave
function at the surface we obtain the quasiclassical spectrum for $|b|\ll
d$ and $|\theta_p|\ll 1$
\begin{eqnarray}
\label{QSpectr}
  \nonumber
  \varepsilon(b,\theta_p,k_z)
  &=&-\frac{\omega k_\perp[(1+h)b+2\theta_pd]}{2}\\
  &\pm&\sqrt{\frac{(\omega k_\perp)^2(1-h)^2b^2}{4}+J^2}\,.
\end{eqnarray}
This expression for the quasiclassical spectrum can also be applied for
the angular interval $|\theta_p-\pi|\ll 1$ provided we replace the angle
$\theta_p\rightarrow\pi-\theta_p$.

\begin{figure}[htb]
\centerline{\includegraphics[width=1.0\linewidth]{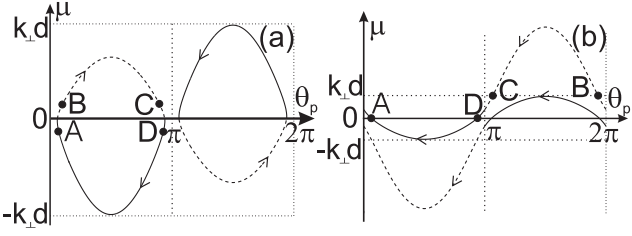}}
  \caption{\label{branches}
The schematic plot of isoenergetic lines $\mu(\theta_p)$ corresponding to
the energy $\varepsilon=0$ for a vortex near the parabolic boundary
positioned at $d>|F|$: $F<0$, $h<0$ (a), $F>0,h>0$ (b). The solid lines
correspond to the trajectories passing close to the vortex core. The dash
lines represent the mapping of solid lines due to the normal reflection of
trajectories at the boundary.}
\end{figure}
 %
The Eqs. (\ref{QSpectrIv}), (\ref{QSpectr}) allow us to determine the form
of isoenergetic lines $\mu(\theta_p)=-k_\perp b(\theta_p)$. Generally, one
can distinguish two types of the isoenergetic lines behavior near the
points $\theta_p=\pi n$ according to the sign of the parameter
$h=-(1+d/F)$. If $h<0$ (i.e., if $F>0$ or $F<0$ and $d<|F|$) there appears
a prohibited angular domain with the width
$\delta\theta_p=\tilde{p}/\sqrt{|\rho|}$ [Fig. \ref{branches}(a)], where
\begin{eqnarray}
\label{p+rho-def}
   \tilde{p}=\frac{J}{\omega\sqrt{|h\rho|}}\,,\\
   \nonumber
   \rho=k_\perp d\frac{F+d/2}{F+d}\,.
\end{eqnarray}
On the other hand, for $h>0$ ($F<0$ and $d>|F|$) there appears a gap
between isoenergetic lines and the prohibited angular domain disappears
[Fig. \ref{branches}(b)].

Solid curves in Fig. \ref{branches} correspond to the trajectories, which
always pass through the vortex core. The reflection of trajectories at the
boundary determines the mapping of a solid curve on the dash curve. To
determine the spectrum we should apply the Bohr--Sommerfeld quantization
rule to the closed path in $(\mu,\theta_p)$--space corresponding to the
precession of the trajectory around the vortex core. Let us study the
formation of such closed path. We start from the consideration of the
trajectory on the solid line, precessing along the orbit in the direction
showed by the arrow in Fig. \ref{branches}. Starting from the point $D$
the trajectory precesses along the solid curve to the point $A$, while the
trajectory on the dash curve moves from the point $B$ towards the point
$C$. As the incident trajectory goes from the point $A$ on the solid curve
to the point $B$ on the dash curve, the corresponding reflected trajectory
goes from the point $C$ on the dash curve to the point $D$ on the solid
curve. Thus, the trajectory is reflected from the boundary and appears at
the point $D$ again, and therefore the quasiclassical orbit in Fig.
\ref{branches} forms a closed path, since the quasiparticle states in the
points $A$ and $D$ are identical. Now we can apply the Bohr--Sommerfeld
quantization rule for the closed orbits in Fig. \ref{branches}:
$S(\varepsilon,k_z)=2\pi n$, where $n$ is integer and $S(\varepsilon,k_z)$
is the area under the solid curves $\mu(\theta_p)$ in Fig. \ref{branches},
taken in the angle domains $0<\theta_p<\pi$ and $\pi<\theta_p<2\pi$.
Evaluating the integral we obtain:
\begin{equation}
\label{area1}
  S(\varepsilon,k_z)=-\pi\frac{\varepsilon}{\omega_s}\pm\left[2k_\perp d
  +{\rm sgn}(h)\tilde{p}^2
  \ln\left(\theta_{th}\sqrt{\frac{|\rho|}{\tilde{p}^2}}
  \right)\right],
\end{equation}
where the interlevel spacing is
\begin{equation}
\label{omegas}
  \omega_s=\omega\left(1+{\rm sgn}(h)\frac{1}{\pi\theta_{th}}
  \frac{\tilde{p}^2}{k_\perp d}\right)^{-1}\,.
\end{equation}
The threshold angle $\theta_{th}$ is of the order of $\xi/(2d)$.

With increasing of the distance from the vortex to the surface the
splitting of the isoenergetic lines tends to zero. According to the
arguments presented in Sec. \ref{SEC:picture} the probability of the
tunneling between different quasiclassical orbits can be estimated as
follows [see Eq. (\ref{LZ1})]:
$W\sim\exp(-2\pi(\delta\theta_p/\Delta\theta_p)^2),$ where
$\delta\theta_p=\tilde{p}/\sqrt{|\rho|}$ and $\Delta\theta_p\sim
1/\sqrt{|\rho|}$. When the splitting is so small that $W\sim 1$ (if
$\tilde{p}\ll1$), the above consideration becomes insufficient and we
should take account of the Landau--Zehner transitions between the
quasiclassical orbits.

To study this limit we employ the approach developed in the Sec.
\ref{SEC:moleculeSub:beyond}. The wave functions should vanish at the
sample surface:
\begin{equation}
\label{bc}
  \int\limits_{0}^{2\pi}e^{ik_\perp R(\theta)\cos(\theta_p-\theta)}
  \hat\psi(R(\theta)\cos(\theta_p-\theta),\theta_p)\,d\theta_p=0\,.
\end{equation}
Using the same technique as in Sec.~\ref{SEC:moleculeSub:beyond} we
obtain:
\begin{eqnarray}
\label{sm1}
  \nonumber
  \left(i\omega\frac{\partial}{\partial\theta}-\varepsilon\right)
  C_1(\theta)&=&e^{i\rho\theta^2-2ik_\perp d}
  \frac{J}{\sqrt{h}}C_2(\theta)\,,\\
  \left(i\omega\frac{\partial}{\partial\theta}
  -\frac{\varepsilon}{h}\right)
  C_2(\theta)&=&e^{-i\rho\theta^2+2ik_\perp d}
  \frac{J}{\sqrt{h}}C_1(\theta)\,,
\end{eqnarray}
where $C_1(\theta)=C(\theta)$ and $C_2(\theta)=iD(\theta/h)$. The boundary
conditions for $C_{1,2}(\theta)$ are:
\begin{eqnarray}
\label{bcSurf}
  \nonumber
  iC_2(h\pi/2)&=&C_1(-\pi/2)\,.\\
  iC_1(\pi/2)&=&C_2(-h\pi/2)\,.
\end{eqnarray}
It is important to notice that Eqs. (\ref{sm1}) are valid until $k_\perp
d|h|\gg 1$. Since the case $h=0$ corresponds to the vortex positioned at
the focal point  of a concave surface, the above condition means that the
vortex distance from the focal point is much larger than the atomic scale.
In case when the vortex is situated at the center of the surface curvature
($d=-2F$, $h=1$ and $\rho=0$) the Eqs. (\ref{sm1})  appear to be very
similar to the Eq. (15) in Ref. \onlinecite{PRB-2007} obtained for a
vortex at the center of a superconducting disc. The only difference is
caused by the absence of the quasiparticle scattering at the opposite end
of the trajectory since we assume the half--infinite superconducting
sample. Hereafter considering the concave surface we focus on the case
when the vortex is shifted from the curvature center towards the boundary
at the distance exceeding the atomic length scale: $|\rho|\gg 1$. Then, at
large angles $\theta\gg J/(\omega \sqrt{|h |}\rho)$ the rapidly
oscillating right hand side in Eq. (\ref{sm1}) can be neglected and we
obtain the solution corresponding to the case of a single vortex in a bulk
superconductor:
\begin{equation}
\label{iVortB}
  {\bf C}=(C_1,C_2)=(a_1 e^{-i\varepsilon\theta/\omega},
  a_2e^{-i\varepsilon\theta/\omega h})\,,
\end{equation}
where $a_1$, $a_2$ are arbitrary constants. In order to obtain the
transfer matrix $\hat X$ in the equation ${\bf C}(\theta_{th})=\hat X{\bf
C}(-\theta_{th})$ we need to consider the domain of small angles.
Introducing new functions
\begin{eqnarray}
  \nonumber
  &B_1(\theta)&=C_1(\theta)e^{ik_\perp d}
  e^{-i\rho\theta^2/2-i\varepsilon_1\theta/\omega}\,,\\
  \nonumber
  &B_2(\theta)&=-{\rm sgn}(\rho)C_2(\theta)
  e^{-ik_\perp
  d}e^{i\rho\theta^2/2-i\varepsilon_1\theta/\omega}\,,
\end{eqnarray}
where $\varepsilon_1=\varepsilon(h+1)/(2h)$, and a new coordinate
$\tilde{\theta}=\sqrt{-\rho}(\theta+\theta_0)$ with
$\theta_0=\varepsilon(h-1)/(2\rho h\omega)$, we obtain the system of
equations for $B_1(\tilde{\theta})$, $B_2(\tilde{\theta})$. This system
coincides with the Eqs. (\ref{TM1}), though for $h<0$ the coordinate $x$
is imaginary and Eqs. (\ref{TM1}) should be solved along the imaginary
axis. Using the solution of Eqs. (\ref{TM1}) (see Appendix
\ref{TransferMat}), we obtain the transfer matrix for $h>0$:
\begin{equation}
\label{TransM+}
  \hat X=e^{-\pi\tilde{p}^2/2}\hat I-i
  \left(\hat\sigma_x{\rm Im}\tau_1+\hat\sigma_y{\rm Re}\tau_1\right)\,,
\end{equation}
where $\hat I$ is the unity matrix,
\begin{eqnarray}
  \nonumber
  \tau_1=\sqrt{2\sinh(\pi\tilde{p}^2/2)}e^{-\pi\tilde{p}^2/4}e^{i\chi_1}\,,
\end{eqnarray}
and $\chi_1=2k_\perp
d+\tilde{p}^2\ln|\tilde{\theta}^*|+\arg(\Gamma(1-i\tilde{p}^2/2))+\pi/4$.
Analogously, for $h<0$ we get:
\begin{equation}
\label{TransM-}
  \hat X=e^{\pi\tilde{p}^2/2}\hat I-
  \left(\hat\sigma_y{\rm Re}\tau_2+\hat\sigma_x{\rm Im}\tau_2\right)
\end{equation}
where
\begin{eqnarray}
  \nonumber
  \tau_2=\sqrt{2\sinh(\pi\tilde{p}^2/2)}e^{\pi\tilde{p}^2/4}e^{i\chi_2}
\end{eqnarray}
and $\chi_2=2k_\perp
d-\tilde{p}^2\ln|\tilde{\theta}^*|-\arg(\Gamma(1-i\tilde{p}^2/2))+\pi/4$.
We denote here
$\tilde{\theta}^*=\sqrt{|\rho|}(\theta_{th}+\theta_0)$.

\begin{figure}[htb]
\centerline{\includegraphics[width=1.0\linewidth]{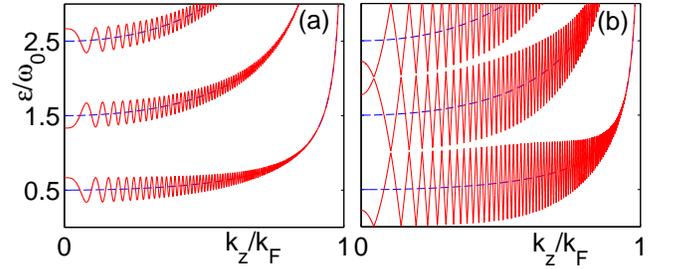}}
  \caption{\label{surf}
The quasiparticle spectrum for the vortex near the flat surface for
$d=2.5\xi$ (a), $d=1.75\xi$ (b). The CdGM spectrum is shown by the dash
lines. The vortex core profile for a single vortex is approximated by Eq.
(\ref{Clem-profile}) with $\xi_v=\xi$, $k_F\xi=200$.}
\end{figure}

The quasiparticle spectrum is obtained by matching the solutions in
different angular domains using the transfer matrices (\ref{TransM+}),
(\ref{TransM-}) and imposing the boundary conditions (\ref{bcSurf}):
\begin{equation}
\label{spSurf}
  \cos\left(\frac{\pi\varepsilon}{\omega}\right)
  =-{\rm sgn}(h)
  \sqrt{2\sinh\left(\frac{\pi\tilde{p}^2}{2}\right)}
  e^{-\pi\tilde{p}^2/4}\coprod\sin\chi\,,
\end{equation}
where
\begin{eqnarray}
\label{xi}
  \nonumber
  \chi&=&2k_\perp d+{\rm sgn}(h)\left[
  \tilde{p}^2\ln\left|\sqrt{|\rho|}(\theta_{th}+\theta_0)\right|\right.\\
  &+&\left.\arg\left(\Gamma\left(1-\frac{i\tilde{p}^2}{2}\right)\right)
  +\frac{\pi}{4}\right].
\end{eqnarray}
The energy $\varepsilon$ enters both the left-- and right-- hand sides
[via $\theta_0=\varepsilon(h-1)/(2\rho h\omega)$] of the Eq.
(\ref{spSurf}). For small energy ($|\varepsilon|\ll\Delta_0$) we have
$\theta_{th}\gg\theta_0$ and the logarithm in Eq. (\ref{xi}) can be
expanded as follows:
$\ln(|\theta_{th}+\theta_0|\sqrt{|\rho|})\approx\ln(\theta_{th}\sqrt{|\rho|})-\theta_0/\theta_{th}$.
The Landau--Zehner transitions between quasiclassical orbits are important
for $\tilde{p}\lesssim 1$ and, therefore, in this limit the
energy--dependent term in $\chi$ can be always neglected:
$\tilde{p}^2\theta_0/\theta_{th}\ll 1$. In the opposite limit the
probability of Landau--Zehner transitions  vanishes and Eq. (\ref{spSurf})
yields the spectrum in the form:
\begin{equation}
\label{spSurfLarge1}
  \varepsilon=\omega\left[2n\pm\frac{2k_\perp d+{\rm sgn}(h)\tilde{p}^2
  \ln\left(\theta_{th}\sqrt{|\rho|/\tilde{p}^2}\right)}{\pi}\right]\,.
\end{equation}
Here $n$ is integer and we have used the asymptotic formula for the
argument of gamma function at $\tilde{p}\gg 1$:
$\arg(\Gamma(1-i\tilde{p}^2/2))\approx-(\tilde{p}^2/2)[\ln(\tilde{p}^2/2)-1]$.
This result coincides with the spectrum obtained from the Bohr--Sommerfeld
quantization rule $S(\varepsilon,k_z)=2\pi n$, where $S(\varepsilon,k_z)$
is given by Eq. (\ref{area1}). The general scenario of spectrum
transformation given by the Eq. (\ref{spSurf}) is shown in Fig.
\ref{surf}.
 %

Comparing Figs. \ref{surf} and \ref{FigSpectrVV} one can see that the
spectrum of a vortex approaching the surface is analogous to the spectrum
of a two--vortex system where the part of energy branches [corresponding
to the upper or lower sign in Eq. (\ref{spectrumVV})] is omitted. To
clarify this result we note that the spectrum transformation in these two
situations is bringing about by the coupling of trajectories with opposite
momentum directions, although the origin of coupling is different. It is
caused by the normal reflection from the boundary when vortex is situated
near the sample surface. For two--vortex or vortex--antivortex systems the
coupling occurs due to the intervortex tunneling and precession of
trajectories around the different vortex cores.  As we have noticed above
in the special case of a vortex near a flat boundary the spectrum exactly
coincides with the spectrum of vortex--antivortex system if we omit the
energy levels corresponding to the even wave functions satisfying
$\partial\hat\Psi(0,y)/\partial x=0$.

\section{Density of states}
\label{SEC:DOS}

Using the results of the analysis of quasiparticle spectrum transformation
under the magnetic field change, we study the corresponding DOS
modification. The calculation of DOS can be done analytically in the
quasiclassical limit, i.e., when the energy can be written as a function
of the classical impact parameter $b=-\mu/k_\perp$, the trajectory
orientation angle $\theta_p$ and the momentum projection $k_z$:
$\varepsilon=\varepsilon(\mu/k_\perp,\theta_p,k_z)$. Solving this equation
for $b$ we find a set of isoenergetic curves
$\mu_i(\theta_p,\varepsilon,k_z)$. Hereafter we focus on the consideration
of DOS contribution coming from a single isoenergetic curve implying that
the summation over the index $i$ enumerating different curves should be
done.

For each isoenergetic curve the set of energy bands $\varepsilon_n(k_{z})$
can be found from the Bohr--Sommerfeld quantization rule:
\begin{equation}
\label{BSdos}
  S(\varepsilon_n,k_{z})=2\pi n,
\end{equation}
where $S(\varepsilon,k_z)$ is the area under the curve
$\mu(\theta_p,\varepsilon,k_z)$. The density of states (per unit length
and per spin projection) is then given by a standard expression:

\begin{eqnarray}
  \nonumber
  \nu(\varepsilon)&=&\sum_n\int\limits_{-k_F}^{+k_F}\frac{dk_z}{2\pi}
  \delta(\varepsilon-\varepsilon_n(k_z))\\
  \label{DOS0}
  &=&\frac{1}{\pi}\sum_n\left|\frac{\partial\varepsilon_n(k_z=q_{n}(\varepsilon))}
  {\partial k_z}\right|^{-1} \ ,
\end{eqnarray}
where the energy spectra $\varepsilon_n (k_z)$ are even functions of
momentum due to the symmetry of the BdG equations with respect to the
$z$--axis inversion and $q_n(\varepsilon)$ is a set of positive momenta
satisfying the equation $\varepsilon_n(q_n)= \varepsilon$. We also assume
here that $|S(\varepsilon,k_z)|$ is a monotonic function of $k_z>0$
reaching the maximal value at $k_z=0$. This condition guarantees that we
get a single positive $q_n$ root for each energy branch. Such assumption
appears to be justified for particular spectrum examples considered above.

Considering the differential of the function $S(\varepsilon, k_z)$ for a
fixed $n$ index we find a simple identity
\begin{eqnarray}
  \nonumber
  \frac{\partial\varepsilon_n(k_{z})}{\partial k_z}
  =-\frac{\partial S/\partial k_z}{\partial S/\partial\varepsilon} \ ,
\end{eqnarray}
which allows us to evaluate the derivatives in Eq. (\ref{DOS0}). As a
first step we neglect the discreteness of the energy spectrum and replace
the sum over $n$ in Eq. (\ref{DOS0}) by the corresponding integral. Taking
a fixed energy in Eq. (\ref{BSdos}) one can transform the differential
$dn$ as follows:
\begin{eqnarray}
  \nonumber
 dn=\frac{1}{2\pi}\frac{\partial S}{\partial k_z}dk_z \ .
\end{eqnarray}
Finally the expression for DOS reads:
\begin{equation}
\label{DOSgen}
  \nu(\varepsilon)=\frac{1}{4\pi^2}\int_{-k_F}^{k_F}
  \left|\frac{\partial S}{\partial\varepsilon}\right|\,dk_z\,.
\end{equation}
Taking the spectrum of a singly--quantized vortex as an example we put
$\mu=-\varepsilon/\omega$ and obtain:
$|S(\varepsilon)|=2\pi\varepsilon/\omega$,
\begin{equation}
\label{nu0}
  \nu(\varepsilon)=\nu_0=\frac{1}{2\pi}\int_{-k_F}^{k_F}\frac{dk_z}{\omega}
  =\frac{k_F}{4\omega_0}\,.
\end{equation}
For a doubly--quantized vortex with the spectrum (\ref{Volovik-spectr})
and $\mu_{1,2}=-\varepsilon/\omega\pm\mu^*$ we get
$\nu(\varepsilon)=k_F/(2\omega_0)=2\nu_0$.

Now we proceed with the calculation of  DOS for a two--vortex system and
vortex near the boundary. For the two--vortex system the quasiclassical
expression for the area under isoenergetic lines has the form
(\ref{S-well-separated}). Therefore, $|\partial
S_{1,2}/\partial\varepsilon|=2\pi/\omega$ and does not depend on the
distance between vortices. Thus, the DOS of the two--vortex system is a
conserved quantity which does not depend on the intervortex distance:
$\nu(\varepsilon)=k_F/2\omega_0=2\nu_0$.

Considering the DOS for a vortex near the boundary of superconductor we
use Eq. (\ref{area1}) and obtain: $|\partial
S_{1,2}/\partial\varepsilon|=2\pi/\omega_s$. The important point is that
$\omega_s$ depends now on the distance from the vortex to the boundary and
on the characteristics of the surface (i.e., the focal distance $F$). The
resulting low--energy DOS can be written as follows: $\nu=\nu_0+{\rm
sgn}(h)\delta\nu$, where
\begin{equation}
\label{Dos1}
  \delta\nu/\nu_0=\frac{2\omega_0}{\pi^2\theta_{th}k_Fd}
  \int_{-k_F}^{k_F}\frac{\tilde{p}^2}{\omega k_\perp}dk_z\,,
\end{equation}
where
 $$
 \tilde{p}=\frac{\Delta_0}{\omega}\frac{e^{-K_0(d)}}{\Lambda\sqrt{k_\perp d|1+d/2F|}}.
 $$
 The ratio $\delta\nu/\nu_0$ is a monotonically decreasing
function of the vortex distance to the surface $d$ and at
$d\gg\xi$ it can be evaluated as follows: $ \delta\nu/\nu_0\sim
(\xi/d)\exp(-4d/\xi)$. Therefore, if $h<0$ ($h>0$), then the DOS
is reduced (increased) as the vortex approaches the surface. If
the vortex is very close to the surface ($d< |F|$), then $h$ is
always negative, therefore the DOS is suppressed. For example, at
$d=\xi$ the correction of DOS $\delta\nu$ is of the order of
$0.1\nu_0$. This effect can be interpreted as the disappearance of
the anomalous spectrum branch occurring as the vortex approaches
the boundary and finally leaves the sample. The DOS reduction is
the direct consequence of the increase of interlevel distance
$\omega_s$ (\ref{omegas}) in the vortex spectrum due to the
appearance of the prohibited domain of trajectory orientation
angles shown in Fig. \ref{branches}(a). The decrease of the
distance $d$ should result in shrinking of the quasiclassical
orbits in $(\mu,\theta_p)$--space and the DOS suppression till to
the complete disappearance of the anomalous spectrum branch at the
moment of vortex exit.

\begin{figure}[htb]
\centerline{\includegraphics[width=1.0\linewidth]{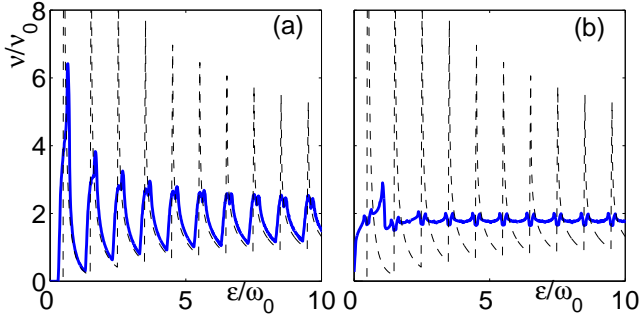}}
  \caption{\label{DOSvv}
The density of states for a two--vortex system for $a=5\xi$ (a),
$a=3.5\xi$ (b). The doubled CdGM DOS is shown by the dash lines. The
vortex core profile for a single vortex is approximated by Eq.
(\ref{Clem-profile}) with $\xi_v=\xi$, $k_F\xi=200$.}
\end{figure}

\begin{figure}[htb]
\centerline{\includegraphics[width=1.0\linewidth]{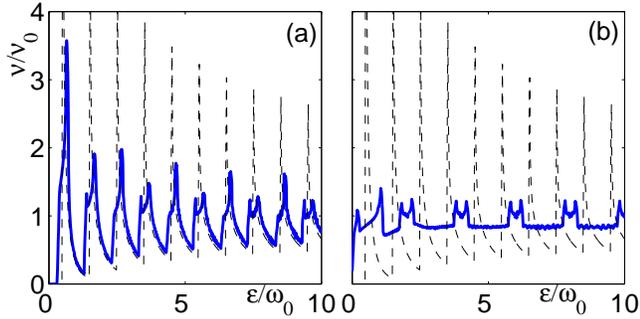}}
  \caption{\label{SurfDOS}
The density of states for a vortex near the flat surface for $d=2.5\xi$
(a), $d=1.75\xi$ (b). The CdGM DOS is shown by the dash lines. The vortex
core profile for a single vortex is approximated by Eq.
(\ref{Clem-profile}) with $\xi_v=\xi$, $k_F\xi=200$.}
\end{figure}

If the spectrum discreteness can not be neglected, the above
quasiclassical calculation of DOS becomes insufficient. In this case, the
rigorous calculation of DOS on the basis of the expression for the
quantized spectrum of the two--vortex system (\ref{spectrumVV}) and the
vortex near the boundary (\ref{spSurf}) can be done  numerically. The
results are shown in the Figs. \ref{DOSvv} and \ref{SurfDOS}. To avoid the
singularities the DOS is averaged over the small energy interval
$0.1\omega_0$. In real experimental conditions such smearing of DOS can be
caused, e.g., by finite temperature or scattering effects.
 %
 %
 %

The DOS of the two--vortex system (Fig. \ref{DOSvv}) in the case $a<a_c$
consists of two sets of small peaks shifted by the value
$\omega_0(2\chi-[2\chi])$ with background level of $2\nu_0$ [Fig.
\ref{DOSvv}(b)]. Here square brackets denote the integer part and $\chi$
is given by the Eq. (\ref{chi-spectrum}). These peaks are van Hove
singularities corresponding to the extrema of the spectrum branches at
$k_z=0$ [Fig. \ref{FigSpectrVV}(b)]. As the distance between vortices
increases the DOS tends to the doubled value of CdGM DOS of the isolated
vortex, shown by the dash lines [Fig. \ref{DOSvv}(a)].

The expression for the spectrum Eq. (\ref{spSurf}) of a vortex near the
boundary is analogous to the spectrum of the two--vortex system, where
part of the  branches, corresponding to the upper or lower sign in Eq.
(\ref{spectrumVV}) is omitted. The function $\nu (\varepsilon)$ for the
vortex near the flat boundary is shown in Fig. \ref{SurfDOS}. One can see
that similarly to the case of a vortex pair the DOS  reveals two sets of
peaks, but the energy scale of DOS oscillations is now larger than
$\omega_0$.

\section{Heat conductance}
\label{SEC:heat-cond}

In this section we calculate the heat conductance of vortex states in a
mesoscopic superconductor focusing on the low temperature limit
$T\ll\Delta_0$ when the transport is dominated by the contribution of
subgap levels. We consider the ballistic regime and neglect the scattering
effects on the boundaries between superconductor and normal metal leads.
The expression for the thermal conductance reads \cite{PRB-2007}:
\begin{equation}
\label{Thcond0}
  \kappa=\frac{1}{4\pi\hbar T^2}\sum_{n}\int_0^{k_F }
  \frac{\varepsilon_n^2}{\cosh^2(\varepsilon_n/2T)}
  \left|\frac{\partial\varepsilon_n}{\partial k_z}\right|dk_z\,.
\end{equation}
Let us introduce the function $N(\varepsilon)$ giving the number of energy
branches crossing a certain energy level $\varepsilon$:
\begin{equation}\label{Ne}
  N(\varepsilon)=\sum_n
  \int\limits_{0}^{k_F} d k_z \delta (\varepsilon-\varepsilon_n(k_z))
  \left|\frac{\partial\varepsilon_n}{\partial k_z}\right|.
\end{equation}
Then the expression (\ref{Thcond0}) can be rewritten as follows:
\begin{equation}
\label{Thcond1}
  \kappa=\frac{1}{4\pi\hbar T^2}\int_0^{\infty}
  \frac{\varepsilon^2N(\varepsilon)}{\cosh^2(\varepsilon/2T)}\,
  d\varepsilon\,.
\end{equation}

In the temperature interval $\omega_0\lesssim T\ll\Delta_0$ the
discreteness of the spectrum can be neglected. In order to evaluate the
number of states $N(\varepsilon)$ in Eq. (\ref{Thcond1}) we use the
quasiclassical theory, assuming that the probability of Landau--Zehner
tunneling between different quasiclassical orbits is small. Generally, the
quasiclassical theory is valid only within the momentum interval
$k_z<k^*_z$, otherwise the interband Landau--Zener transitions can not be
neglected. As a result, in Eq. (\ref{Ne}) we should take the upper limit
of integration $k^*_z$ instead of $k_F$. The value of the threshold
momentum $k_\perp^*$ can be estimated from the condition that the
Landau-Zehner tunneling probability $W$ [see Eq. (\ref{LZest1})] is equal
to a certain threshold value  $W_{th}< 1$. Using the Bohr-Sommerfeld
quantization rule (\ref{BSdos}) we find:
\begin{equation}
\label{NofStates}
  N(\varepsilon)=\left|\int_0^{k^*_z}\frac{d S(\varepsilon,k_z)}{dk_z}
  \frac{dk_z}{2\pi}\right|
  =\frac{|S(\varepsilon,0)-S(\varepsilon,k_z^*)|}{2\pi}\,.
\end{equation}
To evaluate the integral (\ref{Thcond1}) we consider the Taylor expansion
$|S(\varepsilon,0)-S(\varepsilon,k_z^*)|=\sum_{n=0}^{\infty}S^{(n)}\varepsilon^{n}/n!$,
where
\begin{equation}
  \nonumber
  S^{(n)}=\left.\frac{d^n}{d\varepsilon^n}
  |S(\varepsilon,0)-S(\varepsilon,k_z^*)|\right|_{\varepsilon=0}\ .
\end{equation}
As a result, we find the expansion of the heat conductance in power series
of $T$:
\begin{equation}
\label{HCgeneral}
  \kappa=\frac{T}{4\pi\hbar}\sum^{\infty}_{n=0}A_nS^{(n)}T^{n}\,,
\end{equation}
where $A_n$ are expressed in terms of Rihmann function $\zeta(n)$:
\begin{equation}
  \nonumber
  A_n=\frac{2(n+2)(n+1)}{\pi}\left[1-2^{-(n+1)}\right]\zeta(n+2)\,.
\end{equation}
Consequently the effective number of conducting modes
$N_v=\kappa/\kappa_0$ is:
\begin{equation}
  \label{CMgeneral}
  N_v=\frac{3}{4\pi^2}\sum^{\infty}_{n=0}A_nS^{(n)}T^{n}\,,
\end{equation}
For a singly--quantized vortex we get
$N(\varepsilon)=\varepsilon/\omega_0$ and
\begin{equation}
  \nonumber
  N_v=\frac{27\zeta(3)}{2\pi^2}\frac{T}{\omega_0}\,,
\end{equation}
which coincides with the expression obtained in Ref.
\onlinecite{PRB-2007}. For a doubly--quantized vortex
$N(\varepsilon)=2\mu^*$, where $\mu^*\sim k_F\xi$ and $N_v=\mu^*$. Now we
proceed with calculation of the heat conductance for the two--vortex
system and for the vortex near the sample surface.

\subsection{Two--vortex system}

\begin{figure}[htb]
\centerline{\includegraphics[width=1.0\linewidth]{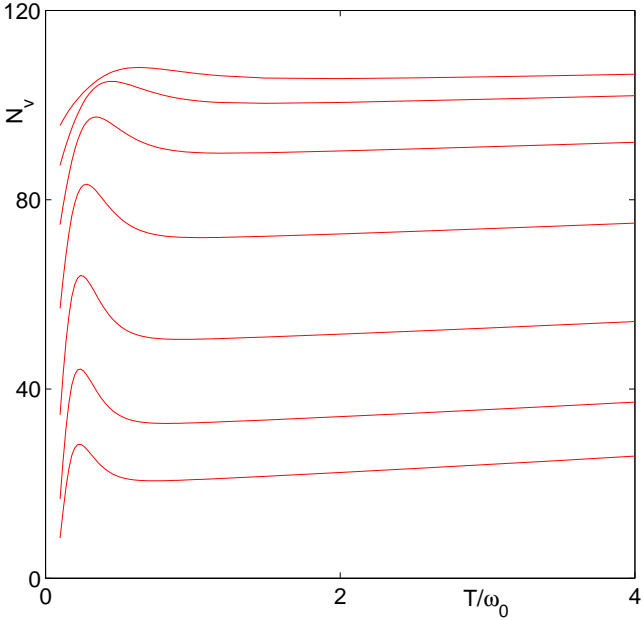}}
  \caption{\label{FigNeff}
Temperature dependence of the number of conducting modes  $N_v$ for a
two--vortex system. Curves are plotted for $a=2\xi$ to $a=5\xi$ with the
step $0.5\xi$ (from top to bottom). The vortex core profile for a single
vortex is approximated by Eq. (\ref{Clem-profile}) with $\xi_v=\xi$,
$k_F\xi=200$. }
\end{figure}

\begin{figure}[htb]
\centerline{\includegraphics[width=1.0 \linewidth]{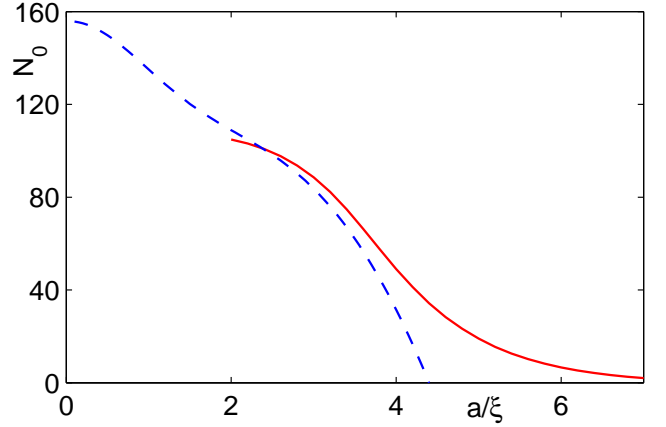}}
  \caption{\label{FigN0}
Residual number of modes as a function of the intervortex distance. Solid
line shows the result of the exact calculation based on Eq.
(\ref{Thcond0}), while dash line is obtained from the analytical
approximate expressions (\ref{CMgeneral}) and (\ref{S-elliptic}).}
\end{figure}

The results of our numerical calculation of the number of conducting modes
as a function of temperature on the basis of Eq. (\ref{Thcond0}) with the
spectrum (\ref{spectrumVV}) are shown in Fig. \ref{FigNeff} .
 %
 %
The suppression of $N_v$ at $T\ll\omega_0$ is caused by the minigap in the
spectrum. One can see that at $T\gtrsim\omega_0$ the function $N_v(T)$
grows linearly with $T$. Extrapolating this linear dependence to $T=0$ we
find the residual number of modes $N_0$, which is plotted by the solid
curve in Fig. \ref{FigN0} as a function of the intervortex distance $a$.
   %

The quasiclassical procedure suggested in the beginning of this section
allows to obtain a good analytical approximation describing the behavior
of the residual number of modes for $a<a_c$. To get such approximation we
evaluate the temperature independent term $N_0=N(0)/2$ in expansion
(\ref{CMgeneral}) which dominates for rather small intervortex distances
$a<a_c$. We use the expression (\ref{NofStates}) for the number of states
$N(\varepsilon)$ where the area confined by the closed orbits in
$(\mu,\theta_p)$--space is given by (\ref{S-elliptic}). The resulting
number of conducting modes vs intervortex distance is shown in Fig.
\ref{FigN0} by the dash curve. Here we choose the threshold probability
$W_{th}=0.54$ to obtain a reasonable fit to the numerical results (solid
curve) at $a\sim 2\xi$. The critical distance $a_c$ is defined by
$N_0(a_c)=0$. The chosen value of the threshold  probability corresponds
to the critical intervortex distance $a_c\simeq4.5\xi$.

For well--separated vortices, i.e. when $a>2\xi$ the splitting of energy
branches is small and one can use an approximate expression (\ref{areaVV})
for the area $S(\varepsilon,k_z)$ to obtain:
\begin{equation}
\label{Neff}
  N_0=\frac{a}{\pi}\left(k_F-k^*_\perp\right)
  +\frac{p^2}{\pi}\ln\left(\theta_{th}\sqrt{k_F a/p^2}\right)\,,
\end{equation}
where
\begin{equation}
\label{k-perp-molec}
  k^*_\perp=k_F\frac{\sqrt{(a/\xi_v)^2+4}-2}{\sqrt{(a_c/\xi_v)^2+4}-2}\,.
\end{equation}
This expression coincides with the estimate (\ref{Threshold}) in the limit
$a\gg\xi_v$.

\subsection{Vortex near the sample boundary}

\begin{figure}[htb]
\centerline{\includegraphics[width=1\linewidth]{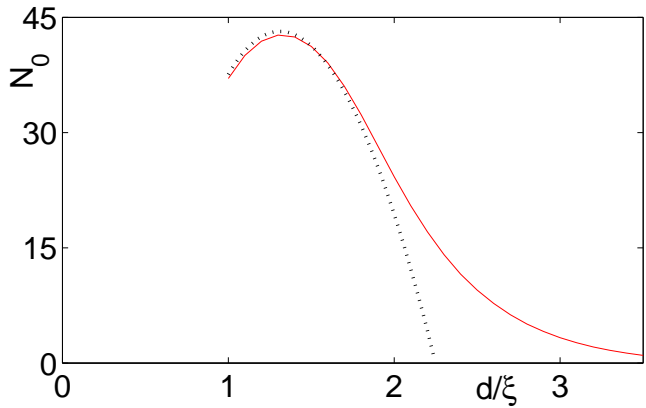}}
  \caption{\label{FigN1}
Residual number of modes $N_0$ as a function of the distance to the
surface $d$. Solid line shows the result of the exact calculation based on
Eq. (\ref{Thcond0}), while dash line is obtained from the analytical
formula (\ref{NofStates2}). The vortex core profile for a single vortex is
approximated by Eq. (\ref{Clem-profile}) with $\xi_v=\xi$, $k_F\xi=200$.}
\end{figure}

The calculation of the number of conducting modes for the vortex near the
boundary can be carried out similarly to the above analysis of the
two--vortex system. We restrict ourselves to the case when the vortex is
situated not very close to the boundary: $d\gtrsim\xi$, when we can
neglect the distortion of vortex core profile due to boundary effects.
Taking for example a flat surface ($h=-1$, $F=\infty$) we use the Eqs.
(\ref{Thcond0}), (\ref{spSurf}) and calculate the residual number of
conducting modes $N_0$ obtained from the extrapolation of the linear parts
of $N_v(T)$ dependencies to $T=0$. In Fig. \ref{FigN1} we plot the
resulting dependence $N_0(d)$ (solid curve).
 %
 %

The estimate of the residual number of conducting modes within the
quasiclassical theory yields the following result:
\begin{equation}
\label{NofStates2}
  N_0=\frac{d}{\pi}(k_F-k^*_\perp)+{\rm sgn}(\kappa)
  \frac{\tilde{p}^2}{2\pi}
  \ln\left(\theta_{th}\sqrt{|\rho|/\tilde{p}^2}\right)\,,
\end{equation}
where
\begin{equation}
\label{k-perp-surf}
  k^*_\perp=k_F\frac{\sqrt{(d/\xi_v)^2+1}-1}{\sqrt{(d_c/\xi_v)^2+1}-1}
\end{equation}
for the particular vortex core model (\ref{Clem-profile}). Taking
$d_c\simeq 2.25\xi$ we plot this approximate expression (\ref{NofStates2})
for $N_0(d)$ in Fig. \ref{FigN1} (dash curve).

One can see that the normal reflection at the surface leads to the
essential increase in the residual number of conducting modes with the
decrease in the distance $d$. This effect is a consequence of the minigap
suppression. The nonmonotonic behavior of $N_0(d)$ at the distances
$d\sim\xi$ has the same origin as the decrease of the zero--energy DOS,
which occurs at these distances. Namely, the residual number of conducting
modes $N_0(d)=N_v(d, \rightarrow 0)$ is defined by the number of states at
the zero energy level $N(0)$ [see Eq. (\ref{HCgeneral})] which is
proportional to the area enclosed by the quasiclassical orbits in
$(\mu,\theta_p)$--space. Therefore the shrinking of closed quasiclassical
orbits at $d\sim\xi$ results in the  decrease in the $N_0(d)$ value. At
small distances to the surface $d<\xi_v$ our approach does not work, but
it is natural to expect further suppression of the subgap conducting modes
number down to zero which accompanies the vortex exit from the sample.

\subsection{Magnetic field dependence of thermal conductance}

\begin{figure}[htb]
\centerline{\includegraphics[width=0.8\linewidth]{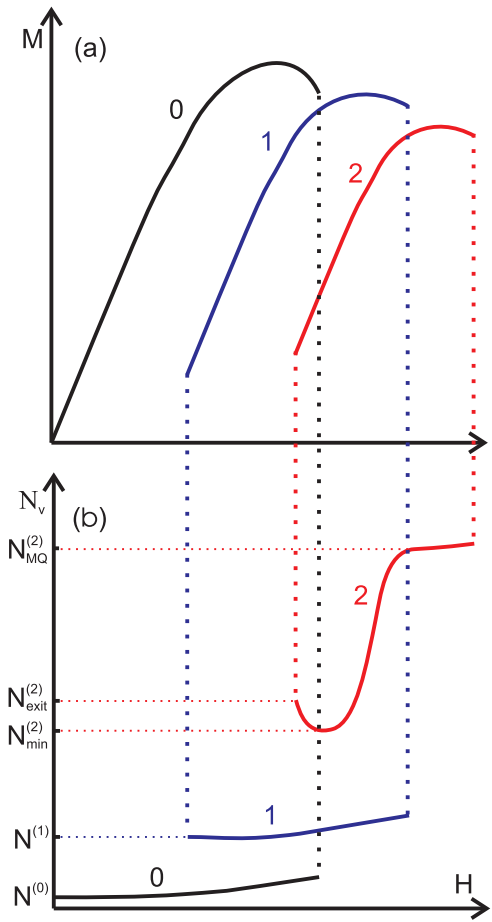}}
  \caption{\label{kappa-H}
Schematic plots of magnetization (a) and effective number of modes
contributing to the heat conductance (b) in a mesoscopic sample vs applied
magnetic field. Different branches correspond to the states with different
number of vortices trapped in the sample.}
\end{figure}

To illustrate our analysis of the heat transport in the vortex state it is
useful to consider the magnetic field dependence of the heat conductance
caused by the transformation of the vortex structure. For this purpose we
plot schematic dependencies of the sample magnetization and effective
number of modes at a certain finite temperature $T\gtrsim\omega_0$ in Fig.
\ref{kappa-H}.
 %
 %
Different branches of the magnetization curve shown in Fig.
\ref{kappa-H}(a) correspond to the different number of vortices in the
superconducting sample. Starting from small magnetic fields $H\ll H_{c2}$
the sample is in the Meissner state, i.e. the number of vortices is zero.
The number of conducting modes $N_v=\kappa/\kappa_0$ [see
Fig.\ref{kappa-H}(b)] is determined by quasiparticle states with energies
above the superconducting gap $\Delta_0$ and therefore is exponentially
small $N_v=N^{(0)}\sim (k_FL)^2e^{-\Delta_0/T}$ provided $T\ll T_c$, where
$L$ is a characteristic transverse size of the sample. In increasing
magnetic field the superconducting gap is suppressed leading to a slightly
growing $N_v$. When the magnetic field becomes large enough to introduce a
vortex into the sample the number of conducting modes jumps to the value
$N^{(1)}\sim T/\omega_0$ simultaneously with the vortex entry. This
increase in $N_v$ is caused by the appearance of subgap quasiparticle
states localized within the vortex core\cite{PRB-2007}.  The next jump in
the number of conducting modes occurs together with the second vortex
entry.  If the sample geometry favors the formation of a giant
doubly--quantized vortex, the number of conducting modes rises up to the
value $N^{(2)}_{MQ}\sim (k_F\xi)$. At the interval of magnetic fields
where the giant vortex is stable, $N_v$ is almost constant. The decay of
the giant vortex into singly--quantized vortices in decreasing magnetic
field is accompanied by the decrease in $N_v$ up to the value
$N^{(2)}_{min}\gtrsim 2N^{(1)}$ (see Section \ref{SEC:molecule} for
details). While the distance between the vortices grows, they approach the
sample surface and the number of conducting modes increases again, as it
was shown in Section \ref{SEC:boundary}. This increase is cut off at a
certain value $N^{(2)}_{exit}$ at the field corresponding to the vortex
exit.

\section{Summary}
\label{DiscusSection}

To summarize, we suggest a description of a subgap quasiparticle spectrum
in the multi--vortex state of a mesoscopic superconductor. Considering
multiquantum (giant) vortices we have obtained a general analytical
expression for the quasiparticle spectrum, which is valid for any value of
vorticity and arbitrary vortex core model. Taking the simplest example of
the doubly--quantized vortex we have considered the evolution of the
anomalous spectrum branches which accompanies the splitting of the
doubly--quantized vortex. Considering the limit of well--separated
vortices we have found the spectrum of vortex clusters bonded by the
quasiparticle tunneling and have investigated the crossover to the
Caroli--de Gennes--Matricon spectrum of isolated vortices. We have shown
that the minigap in the quasiparticle spectrum is absent for the
intervortex distances $a<a_c\approx\xi\ln (k_F\xi)$. In mesoscopic
superconductors it is necessary also to take account of the normal
reflection of quasiparticles at the sample surface. We have shown that the
spectrum of a single vortex placed near the parabolic surface is
transformed analogously to the two--vortex system and the minigap in the
spectrum is suppressed when the distance from the vortex to the surface is
less than the critical value: $d<d_c\approx(\xi/2)\ln (k_F\xi)$. When the
distance is of the order of the vortex core size, the interlevel spacing
in the vortex spectrum becomes larger than the CdGM value. This effect
leads to the disappearance of the anomalous spectrum branch when the
vortex approaches the surface.

We have analyzed the quasiparticle density of states and the heat
conductance along the magnetic field, which are determined by the
anomalous branches of the quasiparticle spectrum. At the temperatures
$T\gg\omega_0$ neglecting the discreteness of spectrum we have obtained a
general expression for the DOS and heat conductance through the
characteristics of the quasiclassical orbits in $(\mu,\theta_p)$--space.
Applying the general formula to the vortex pair we have observed a
significant decrease of the heat conductance as a function of the growing
intervortex distance. Even in the limit of the zero intervortex distance,
i.e. for a doubly--quantized vortex, the number of conducting modes
$N_v\sim k_F\xi$ appears to be much less than the value $(k_F\xi)^2$,
which determines the number of conducting modes for a normal metal wire of
the radius $\xi$. At nonzero intervortex distances and in the temperature
region $\omega_0<T\ll\Delta_0$ the effective number of transport modes is
a linear function of temperature: $N_v=N_0+\alpha T$. The splitting of the
doubly--quantized vortex is accompanied by the decrease of the residual
number of modes $N_0$ and at rather large intervortex distances we get the
doubled heat conductance of a single vortex: $N_v\sim T/\omega_0\ll
k_F\xi$.

Also we have shown that the normal reflection at the surface of the sample
leads to a considerable increase in the heat conductance along the
magnetic field when the distance from the vortex to the sample boundary
becomes rather small: $\xi\lesssim d<d_c$. The exit of a vortex from the
sample is accompanied by the disappearance of the anomalous spectrum
branch and, therefore, both the heat conductance and the DOS are
suppressed at $d\lesssim\xi$.

\section{Acknowledgements}
We thank N.~B. Kopnin and G.~E. Volovik for stimulating discussions, I.~A.
Shereshevskii and V.~I. Pozdnyakova for the help with computer codes. This
work was supported, in part, by Russian Foundation for Basic Research, by
Program ``Quantum Macrophysics'' of RAS, and by Russian Science Support
and ``Dynasty'' Foundations.

\appendix
\section{Derivation of Eq. (\ref{r11})}
\label{ScaleTr}

At first let us prove the following formula:
\begin{equation} \label{A1}
  \int_{-\infty}^{\infty}e^{-ik^2/2+ikx}F(k)\,dk
  =\sqrt{2\pi i} \int_{-\infty}^{\infty}e^{i(x-y)^2/2}f(y)\,dy\,,
\end{equation}
where $f(x)$ is a smooth enough function defined at
$-\infty<x<\infty$ and $F(k)=2\pi\int_{-\infty}^{\infty} e^{-
ikx}f(x)dx$. Indeed, the integral in the right hand side of Eq.
(\ref{A1}) can be written as follows:
\begin{equation}
  \nonumber
  \int_{-\infty}^{\infty}e^{i(x-y)^2/2}f(y)\,dy
  =\iint_{-\infty}^{\infty}e^{i(x-y)^2/2+iky}F(k)\,dk\,dy\,.
\end{equation}
Noting that $(x-y)^2/2+ky=(y+k-x)^2/2+kx-k^2/2$ we integrate over
the $y$ variable using the formula
$\int_{-\infty}^{\infty}e^{ix^2}dx=\sqrt{\pi i}$ and obtain
\begin{eqnarray}
  \nonumber
  \iint_{-\infty}^{\infty}e^{i(x-y)^2/2+iky}F(k)\,dk\,dy\\
  \nonumber
  =\frac{1}{\sqrt{2\pi
  i}}\int_{-\infty}^{\infty}e^{-ik^2/2+ikx}F(k)\,dk.
 \end{eqnarray}
Taking Eq. (\ref{r1}) in the form
\begin{eqnarray}
  \label{A2}
  \nonumber
  \left[\int_\infty^\infty e^{i\mu^2/k_\perp a}
  \left(e^{-K_0(a/2)}-i\hat\gamma\hat\sigma_z\right)
  \hat\lambda e^{i\mu\theta_1}c_\mu d\mu\right]=\\
  e^{-2i\varphi}\left[\int_\infty^\infty e^{-i\mu^2/k_\perp a}
  \left(e^{-K_0(a/2)}+i\hat\gamma\hat\sigma_z\right)
  \hat\lambda^* e^{-i\mu\theta_1}c^*_\mu d\mu\right]
  \,,
\end{eqnarray}
we multiply it by $e^{ik_\perp a(\theta-\theta_1)^2/4}$ and
integrate over $\theta_1$. Using Eq. (\ref{A1}) we can transform
the above integrals as follows:
\begin{eqnarray}
  \nonumber
  \iint_{-\infty}^\infty  e^{ik_\perp a(\theta-\theta_1)^2/4}
  e^{i\mu^2/k_\perp a}e^{i\mu\theta_1}c_\mu d\mu d\theta_1=\\
  \nonumber
  \sqrt{\frac{2\pi}{i}} \iint_{-\infty}^\infty e^{i(k_\perp a/4)
  \left[(\theta-\theta_1)^2-(\theta_1-\theta_2)^2\right]}C(\theta_2)
  d\theta_1 d\theta_2=\\
  \nonumber
  =\frac{\sqrt{2/i\pi}}{k_\perp a}C(\theta),\\
  \nonumber
  \iint_{-\infty}^\infty e^{ik_\perp a
  \left[(\theta-\theta_1)^2/4-\theta^2/2\right]}
  e^{-i\mu^2/k_\perp a}e^{-i\mu\theta_1}c^*_\mu d\mu=\\
  \nonumber
  \sqrt{2\pi i}\iint_{-\infty}^\infty e^{ik_\perp a
  \left[(\theta-\theta_1)^2/4+(\theta_1-\theta_2)^2/4-\theta_1^2/2\right]}
  C(\theta_2) d\theta_1 d\theta_2=\\
  \nonumber
  =\frac{\sqrt{2i/\pi}}{k_\perp a} e^{ik_\perp a\theta^2/2}C^*(-\theta)
\end{eqnarray}
Making use of these expressions the derivation of Eq. (\ref{r11})
from Eq. (\ref{A2}) is straightforward.

\section{Transfer matrices}
\label{TransferMat}

Let us consider the following system of equations:
\begin{eqnarray}
\label{2A1}
  \nonumber
  i\frac{\partial}{\partial x}B_1+x B_1=pB_2\,,\\
  i\frac{\partial}{\partial x}B_2-x B_2=pB_1\,,
\end{eqnarray}
where $p>0$ and $x$ is a coordinate along real or imaginary axis.
We start with the case of real $x$. Solutions of Eqs. (\ref{2A1})
can be expressed in terms of the parabolic cylinder functions $D$
(Ref. \onlinecite{Whittaker-Watson-1947}) with arbitrary constants
$d_1$ and $d_2$:
\begin{eqnarray}
\label{2A2}
  \nonumber
  B_1&=&d_1D_{ip^2/2}\left(x\sqrt{\frac{2}{i}}\right)
  +d_2D_{ip^2/2}\left(-x\sqrt{\frac{2}{i}}\right)\,,\\
  B_2&=&\frac{p}{\sqrt{2i}}\left[
  d_1D_{ip^2/2-1}\left(x\sqrt{\frac{2}{i}}\right)\right.\\
  \nonumber
  &-&\left.
  d_2D_{ip^2/2-1}\left(-x\sqrt{\frac{2}{i}}\right)\right]\,.
\end{eqnarray}
The asymptotic expressions for the obtained solutions for
$x\gg\max(1,p)$ are following:
\begin{eqnarray}
  \nonumber
  D_{ip^2/2}\left(x\sqrt{\frac{2}{i}}\right)
  \simeq e^{ix^2/2+i(p^2/2)\ln(\sqrt{2}x)+\pi p^2/8},\\
  \nonumber
  D_{ip^2/2-1}\left(x\sqrt{\frac{2}{i}}\right)\simeq 0,\\
  \nonumber
  D_{ip^2/2}\left(-x\sqrt{\frac{2}{i}}\right)
  \simeq e^{i x^2/2+i(p^2/2)\ln (\sqrt{2}x)-3\pi p^2/8},\\
  \nonumber
  D_{ip^2/2-1}\left(-x\sqrt{\frac{2}{i}}\right)\simeq\sqrt{2\pi}
  \frac{e^{-ix^2/2-i(p^2/2)\ln(\sqrt{2}x)-\pi p^2/8}}{\Gamma(1-ip^2/2)},
\end{eqnarray}
where $\Gamma$ is the gamma function. Then, we find the scattering
matrix $\hat X_1$ coupling the solutions $\hat B=(B_1,B_2)$ at
$x>0$ and $x<0$:
 $$
 \hat B(x>0)= \hat X_1 \hat B(x<0)
 $$
in the following form:
\begin{equation}
  \hat X_1=e^{-\pi p^2/2}\hat I
  +i\left(\hat\sigma_y{\rm Re}\tau_1+\hat\sigma_x{\rm Im}\tau_1\right)\,,
\end{equation}
where $\hat I$ is the unity matrix,
\begin{equation}
  \nonumber
  \tau_1=\sqrt{2\sinh(\pi p^2/2)}e^{-\pi p^2/4}e^{i\chi_1}\,,
\end{equation}
and $\chi_1=x^2+p^2\ln|\sqrt{2}x|+\arg\Gamma(1-p^2/2)+\pi/4$.

Next, we consider the case of imaginary coordinate $x$.
Introducing a new variable $y=-ix$ we obtain an analytical
continuation of the solutions (\ref{2A2}):
\begin{eqnarray}
  \nonumber
  B_1&=&d_1D_{ip^2/2}(y\sqrt{2i})+d_2D_{ip^2/2}(-y\sqrt{2i})\,,\\
  \nonumber
  B_2&=&\frac{p}{\sqrt{2i}}
  \left[d_1D_{ip^2/2-1}(y\sqrt{2i})-d_2D_{ip^2/2-1}(-y\sqrt{2i})\right]\,.
\end{eqnarray}
The asymptotic expressions for this solutions at $y\gg\max(1,p)$
have the form:
\begin{eqnarray}
  \nonumber
  D_{ip^2/2}(y\sqrt{2i})&\simeq&
  e^{-iy^2/2+i(p^2/2)\ln (\sqrt{2}y)-\pi p^2/8}\,,\\
  \nonumber
  D_{ip^2/2-1}(y\sqrt{2i})&\simeq& 0\,,\\
  \nonumber
  D_{ip^2/2}(-y\sqrt{2i})&\simeq&
  e^{-i y^2/2+i(p^2/2)\ln (\sqrt{2}y)+3\pi p^2/8}\,,\\
  \nonumber
  D_{ip^2/2-1}(-y\sqrt{2i})&\simeq&
  \sqrt{2\pi}\frac{e^{iy^2/2-i(p^2/2)\ln(\sqrt{2}y)+\pi p^2/8}}
  {\Gamma (1-ip^2/2)}\,,
\end{eqnarray}
and the transfer matrix is
\begin{equation}
  \hat X_2=e^{\pi p^2/2}\hat I+\left(\hat\sigma_y{\rm Re}\tau_2+
  \hat\sigma_x{\rm Im}\tau_2\right)\,,
\end{equation}
where
\begin{equation}
  \nonumber
  \tau_2=\sqrt{2\sinh(\pi p^2/2)}e^{\pi p^2/4}e^{i\chi_2}\,,
\end{equation}
and $\chi_2=y^2-p^2\ln|\sqrt{2}y|-\arg \Gamma(1-p^2/2)+\pi/4$.


\begin{thebibliography}{99}
\bibitem{Mesovortices}
G.~Boato, G.~Gallinaro, C.~Rizutto, Solid St. Commun. {\bf 3}, 173 (1965);
D.~S. McLachlan, Solid St. Commun. {\bf 8}, 1589 (1970); L.~F. Chibotaru,
A.~Ceulemans, V.~Bruyndoncx, and V.~V. Moshchalkov, Nature {\bf 408}, 833
(2000).

\bibitem{Meso-Diagram}
V.~A. Schweigert, F.~M. Peeters, and P.~Singha Deo, Phys. Rev. Lett. {\bf
81}, 2783 (1998).

\bibitem{Geim}
A.~K. Geim, S.~V. Dubonos, J.~J. Palacios, I.~V. Grigorieva, M.~Henini,
and J.~J. Schermer, Phys. Rev. Lett. {\bf 85} 1528 (2000).

\bibitem{M-quantum-defect}
V.~V. Schmidt and G.~S. Mkrtchyan, Usp. Fiz. Nauk {\bf 112}, 459 (1974)
[Sov. Phys. Usp. {\bf 17}, 170 (1974)]; G.~R. Berdiyorov, B.~J. Baelus,
M.~V.
Milo\ifmmode\check{s}\else\v{s}\fi{}evi\ifmmode\acute{c}\else\'{c}\fi{},
and F.~M. Peeters, Phys. Rev. B, {\bf 68}, 174521, (2003).

\bibitem{Peeters}
A.~Kanda, B.~J. Baelus, F.~M. Peeters, K.~Kadowaki, and Y.~Ootuka,
Phys. Rev. Lett. {\bf 93}, 257002 (2004); B.~J. Baelus, A.~Kanda,
F.~M. Peeters, Y.~Ootuka and K.~Kadowaki, Phys. Rev. B {\bf 71},
140502 (2005); B.~J. Baelus, A.~Kanda, N.~Shimizu, K.~Tanado,
Y.~Ootuka, K.~Kadowaki, and F.~M. Peeters, Phys. Rev. B {\bf 73},
024514 (2006).

\bibitem{CdGM}
C.~Caroli, P.~G. de Gennes, J.~Matricon, Phys. Lett. {\bf 9}, 307
(1964).

\bibitem{Melnikov-Silaev-2006}
A.~S. Melnikov, M.~A. Silaev, Pis'ma Zh. Eksp. Teor. Fiz. {\bf
83}, 675 (2006) [JETP Lett. {\bf 83}, 578 (2006)].

\bibitem{Volovik-1993}
G.~E. Volovik, Pis'ma Zh. Eksp. Teor. Fiz. {\bf 57}, 233 (1993)
[JETP Lett. {\bf 57}, 244 (1993)].

\bibitem{multi-spectrum-num}
Y.\ Tanaka, A.\ Hasegawa, and H.\ Takayanagi, Solid St. Commun. {\bf 85},
321 (1993); Y.~Tanaka, S.~Kashiwaya, and H.~Takayanagi, Jpn. J. Appl.
Phys. {\bf 34}, 4566 (1995); D.\ Rainer, J.\ A.\ Sauls, and D.\ Waxman,
Phys. Rev. B {\bf 54}, 10094 (1996); S.~M.~M. Virtanen and M.~M. Salomaa,
Phys. Rev. B {\bf 60}, 14581 (1999); M.~Eschrig, D.~Rainer, and J.~A.
Sauls: in {\it Vortices in unconventional superconductors and
superfluids}, ed. R.~P. Huebener, N.~Schopohl and G.~E. Volovik (Springer
Verlag, Berlin, 2001), [preprint cond-mat/0106545]; K.\ Tanaka, I.\ Robel,
and B.\ Janko, Proc. Nat. Acad. Sci. (USA) {\bf 99}, 5233 (2002).

\bibitem{Melnikov-Vinokur-2002}
A.~S. Mel'nikov and V.~M. Vinokur, Nature, {\bf 415}, 60 (2002);
Phys. Rev.~B {\bf 65}, 224514 (2002).

\bibitem{Larkin-Ovchinnikov-1998}
A.~I. Larkin and Yu.~N. Ovchinnikov, Phys. Rev.~B {\bf 57}, 5457
(1998).

\bibitem{dahm} S.~Graser, C.~Iniotakis, T.~Dahm, N.~Schopohl,
Phys. Rev. Lett. {\bf 93}, 247001, (2004).

\bibitem{PRL-2005}
N.~B. Kopnin, A.~S. Mel'nikov, V.~I. Pozdnyakova, D.~A. Ryzhov,
I.~A. Shereshevskii, and V.~M. Vinokur, Phys. Rev. Lett. {\bf 95},
197002 (2005).

\bibitem{PRB-2007}
N.~B. Kopnin, A.~S. Mel'nikov, V.~I. Pozdnyakova, D.~A. Ryzhov,
I.~A. Shereshevskii, and V.~M. Vinokur, Phys. Rev.~B {\bf 75},
024514 (2007).

\bibitem{KMV03}
N.~B. Kopnin, A.~S. Mel'nikov, and V.~M. Vinokur, Phys. Rev. B
{\bf 68}, 054528 (2003).

\bibitem{LL-III}
L.~D. Landau, E.~M. Lifshitz "Quantum mechanics. Non-relativistic
theory", Pergamon Press, 1991.

\bibitem{Kopnin-Volovik-1996}
N.~B. Kopnin and G.~E. Volovik, Pis'ma Zh. Eksp. Teor. Fiz. {\bf
64}, 641 (1996) [JETP Lett. {\bf 64}, 690 (1996)]; N.~B. Kopnin,
Phys.~Rev.~B {\bf 57}, 11775 (1998).



\bibitem{kane}
E.~O. Kane and E.~I. Blount, in {\sl Tunneling Phenomena in
Solids}, Eds. E. Burstein and S. Lundqvist, New York, Plenum
Press, 1969.

\bibitem{Whittaker-Watson-1947}
E.~T. Whittaker and G.~N. Watson, {\it Modern Analysis}
(Cam\-bridge University Press, 1947), Chap. 16.

\end{thebibliography}
\end{document}